\documentclass[prc,twocolumn,floatfix,groupedaddress,nofootinbib,showpacs,preprintnumbers,
amsmath,amssymb,amsfonts,superscriptaddress,widetable] {revtex4-1}
\usepackage{bm}
\usepackage{mathrsfs}
\usepackage{amssymb}
\usepackage{amsmath}
\usepackage{graphicx}
\usepackage{array}
\usepackage[normalem]{ulem}
\usepackage{xcolor}

\begin{document}
\title{Deep Crustal Heating by Neutrinos from the Surface of Accreting Neutron Stars}
\author{F. J. Fattoyev}\email{ffattoye@indiana.edu}
\affiliation{Center for Exploration of Energy and Matter and
                  Department of Physics, Indiana University,
                  Bloomington, IN 47405, USA}
\author{Edward F. Brown}\email{ebrown@pa.msu.edu}
\affiliation{Department of Physics and Astronomy, Michigan State
                  University, 567 Wilson Rd, East Lansing, MI 48864, USA}
\author{Andrew Cumming}\email{andrew.cumming@mcgill.ca}
\affiliation{Department of Physics and McGill Space Institute,
                  McGill University, 3600 rue University, Montreal QC, Canada H3A
                  2T8}
\author{Alex Deibel}\email{adeibel@indiana.edu}
\affiliation{Department of Astronomy, Indiana University,
                  Bloomington, IN 47405, USA}
\author{C. J. Horowitz}\email{horowit@indiana.edu}
\affiliation{Center for Exploration of Energy and Matter and
                  Department of Physics, Indiana University,
                  Bloomington, IN 47405, USA}
\author{Bao-An Li}\email{bao-an.li@tamuc.edu}
\affiliation{Department of Physics and Astronomy, Texas A\&M
                  University-Commerce, Commerce, TX 75429, USA}
\author{Zidu Lin}\email{zidulin@imail.iu.edu}
\affiliation{Center for Exploration of Energy and Matter and
                  Department of Physics, Indiana University,
                  Bloomington, IN 47405, USA}
\date{\today}
\begin{abstract}
We present a new mechanism for deep crustal heating in accreting
neutron stars. Charged pions ($\pi^+$) are produced in nuclear
collisions on the neutron star surface during active accretion and
upon decay they provide a flux of neutrinos into the neutron star
crust. For massive and/or compact neutron stars, neutrinos deposit
$\approx 1\textrm{--} 2 \, \mathrm{MeV}$ of heat per accreted
nucleon into the inner crust. The strength of neutrino heating is
comparable to the previously known sources of deep crustal heating,
such as from pycnonuclear fusion reactions, and is relevant for
studies of cooling neutron stars. We model the thermal evolution of
a transient neutron star in a low-mass X-ray binary, and in the
particular case of the neutron star MXB~1659-29 we show that
additional deep crustal heating requires a higher thermal
conductivity for the neutron star inner crust. A better knowledge of
pion production cross sections near threshold would improve the
accuracy of our predictions.
\end{abstract}

\smallskip
 \maketitle

\section{Introduction}
\label{Introduction} Neutron stars in X-ray binaries accrete matter
from their companion stars. As matter is accreted by the star, the
crust is continually compressed and undergoes a series of
non-equilibrium nuclear reactions such as electron captures, neutron
emissions and pycnonuclear fusion reactions that release $\approx
1\textrm{--}2\, \mathrm{MeV}$ per accreted nucleon
\cite{bisnovatyi1979,sato1979,haensel1990,haensel2003,Haensel:2007wy}.
Energy release mainly occurs in the inner crust at mass densities of
about $10^{12} \textrm{--} 10^{13} \, \mathrm{g \ cm^{-3}}$ and is
referred to as deep crustal heating.
During an accretion outburst, deep crustal heating brings the entire
crust out of thermal equilibrium with the core. When accretion ends
and the neutron star enters quiescence, crust cooling powers an
observable X-ray light curve~\cite{Ushomirsky:2001pd,
Rutledge:2002,shternin2007}. Thermal evolution models of accreting neutron stars
that include deep crustal heating successfully reproduce most
observed quiescent X-ray light curves~\cite{Brown:2009kw}. The
cooling light curves of several sources, however, require an
additional heat deposition in the outer crust during outburst to
reach observed quiescent temperatures~\cite{Cumming:2005kk,
Deibel:2015ola}. The source of extra heating remains unknown, but
must be comparable in strength to the heat release from
non-equilibrium nuclear reactions.

Here we discuss a new source of heating in neutron star crusts from
the decay of charged pions on the neutron star surface. The neutron
star's strong gravity accelerates incoming particles to kinetic
energies of several hundred MeV per nucleon before they reach the
neutron star surface. The accreted matter, usually consisting of
hydrogen or helium, undergoes nuclear collisions with the nuclei on
the neutron star surface. Nuclear collisions produce pions, in
particular $\pi^{+}$, that upon decay emit a flux of neutrinos.
Approximately half of these neutrinos carry their energy into the
crust, where they experience multiple scatterings and are eventually
absorbed in the inner crust. This neutrino heating provides an
additional source for deep crustal heating. In this work we present
the first calculations of deep crustal heating by neutrinos from the
decay of stopped pions on the surface of neutron stars.

The paper is organized as follows. In Section \ref{Formalism} we
develop the formalism required to discuss the energy deposition by
neutrinos from the stopped pion decays. We will first briefly review
the main mechanism of deep crustal heating by neutrinos in
\ref{Mechanism}. Following this discussion, in \ref{NucIntNSE}, we
review the main steps involved in calculating the energy deposited
from pion production on the surface of neutron stars. We close this
Section by discussing in \ref{EnDeposit} the approximate location of
the inner crust where neutrinos will deposit their energies. We then
proceed to Section \ref{Results} to display the results of our
calculations using various equations of state (EOS) as well as pion
production from three possible nuclear reactions. This section ends
with a discussion of the observational implications of deep crustal
heating in understanding cooling light curves of neutron stars in
X-ray transients. Finally, we offer our conclusions in Section
\ref{Conclusions}.

\section{Formalism}
\label{Formalism}

\subsection{The Main Mechanism}
\label{Mechanism} Neutron stars in low-mass X-ray binaries typically
accumulate hydrogen-rich or helium-rich matter from the surface of
their companions in an accretion disk, a rotating disk of matter
formed by accretion around the neutron star. The accumulated matter
is later accreted onto the neutron star surface during an accretion
outburst stage which makes neutron stars as bright X-ray sources.
If the innermost stable orbit lies outside the neutron star, or if
the accretion disk is truncated by the magnetic field of the star,
the final trajectory of the accreted material can be close to
radial, and the matter arrives at the neutron star surface with
close to the free-fall velocity. For such quasi-spherical flows most
of the gravitational potential energy is retained in the incoming
particles in the form of the kinetic energy that particles carry
into the neutron star. During outburst, these incoming particles
collide with the nuclei on the neutron star
surface~\cite{Bildsten:1992ApJ} and can produce pions if the
particle's kinetic energy is above the pion production threshold of
$\approx 290\, \mathrm{MeV}$. Neutral pions decay almost
instantaneously via $\pi^0 \rightarrow 2 \gamma$ releasing their
energy at the surface. Neutrinos from the decay of negative pions
may be strongly suppressed because $\pi^-$ are often absorbed, via
strong interactions, before they can undergo a weak decay.
Positively charged pions slow down and stop near the neutron star
surface and decay into muons and muon neutrinos through $\pi^{+}
\rightarrow \mu^{+}\nu_{\mu}$. This produces monoenergetic muon
neutrinos, $\nu_{\mu}$, of energy $E_{\nu_{\mu}} = 29.8$ MeV. The
anti-muon subsequently decays through $\mu^{+} \rightarrow e^{+}
\nu_{e} \bar{\nu}_{\mu}$ on a muon-decay time scale of $\tau = 2.2
\, \mu s$, with a well-determined neutrino energy
spectrum~\cite{Scholberg:2005qs}. Approximately half of the
neutrinos produced escape the neutron star and the other half move
into the crust carrying a total energy of
\begin{equation}
\label{TotQnu} Q_{\nu} \approx 0.5 \left(E_{\nu_{\mu}} + E_{\nu_{e}}
+ E_{\bar{\nu}_{\mu}} \right) N_{\pi^+} = (50.4\ {\rm MeV})\,
N_{\pi^+} \
\end{equation}
per accreted nucleon, where $N_{\pi^+}$ is the total number of
$\pi^+$'s produced per accreted nucleon. In addition to
gravitational acceleration, accreting particles may undergo
electromagnetic acceleration in the strong electric and magnetic
fields that are likely present. This could significantly increase
pion and neutrino production, but we will explore this in later
work.

\subsection{Pion Production per Accreted Nucleon}
\label{NucIntNSE}

We now calculate the number of charged pions produced from infalling
matter. Assuming that the infalling matter has zero velocity at
infinity (free-falling), we estimate the kinetic energy $T$ of the
accreted matter at the surface of the neutron
star~\cite{Bildsten:1992ApJ} using
\begin{equation}
T  = m_0 c^2 \left(\frac{1}{\sqrt{1 - R_{\rm S}/R}}  -1 \right) \ ,
\end{equation}
where $R_{\rm S} \equiv 2 G M/c^2$ is the Schwarzschild radius,
$m_0$ is the mass of the infalling particle, and $M$ and $R$ are the
neutron star mass and radius, respectively. If the kinetic energy of
the incoming particles is sufficiently large, they will collide with
the nuclei on the surface of the neutron star and can produce pions.
The multiplicity of pion production, defined as the number of pions
produced per collision event, strongly depends on the initial
kinetic energy of the incoming particle as well as on the type of
the target nuclei. Here the target nuclei on the surface of neutron
stars could be composed of any mixtures of light-to-medium nuclei.
Since both incoming protons and $\alpha$-particles are charged
particles, before they undergo a hard nuclear collision, they
partially lose energy due to interaction with atmospheric electrons.
The energy loss of charged particles can be calculated using the
Bethe-Bloch equation
\begin{equation}
-\frac{dE}{dx}  = K \frac{Z_{\rm p}^2}{\beta^2} \frac{Z_{\rm
t}}{A_{\rm t}}\left(\frac{1}{2}\ln\frac{2m_e c^2\beta^2 \gamma^2
T_{\rm max}}{I^2} - \beta^2 - \frac{\delta}{2} \right) \ ,
\end{equation}
where $K  \approx 0.307075$ MeV mol$^{-1}$ cm$^2$, $Z_{\rm p}$ is
the charge number of the incident particle (projectile), $Z_{\rm t}$
is the atomic number of the target, $A_{\rm t}$ is the atomic mass
of the target in g mol$^{-1}$, $\beta =v/c$, $\gamma =
1/\sqrt{1-\beta^2}$ is the relativistic Lorentz factor, $I$ is the
mean excitation energy, and $\delta$ is the density effect
correction to ionization energy loss which is negligible for
energies under consideration. Here
\begin{equation}
T_{\rm max} = \frac{2m_e c^2\beta^2 \gamma^2}{1+2\gamma m_e/m_0
+(m_e/m_0)^2} \ ,
\end{equation}
is the maximum kinetic energy which can be imparted to a free
electron in a single collision. A complete description of the
electronic energy loss by heavy particles can be found in Chapter 32
of Ref.~\cite{Amsler:2008zzb}. A similar study of the incident-beam
particles deceleration through repeated Coulomb scatters from
atmospheric electrons was also carried out in
Ref.~\cite{Bildsten:1992ApJ}. The energy of the particle that
undergoes a hard nuclear collision is therefore
\begin{equation}
\label{EqnEx} E_{\rm f}(x) \approx E_{\rm i}(x) + \lambda
\frac{dE}{dx} \ ,
\end{equation}
where $\lambda = 1/n \sigma$ is the strong interaction mean free
path, $n$ is the number density of scattering centers, and $\sigma$
is the strong collision cross section. Note that the energy loss
$\frac{dE}{dx}$ depends on the initial beam energy $E_{\rm i}(x)$
through Lorentz parameters. Therefore, Eqn. (\ref{EqnEx}) takes an
exact form if one replaces $\lambda$ with $\Delta x = \lambda/N$,
where $N \gg 1$, and solves the recurrent relation
\begin{equation}
\label{EqnEx2} E(x+\Delta x) = E(x) + \Delta x \frac{dE}{dx} \ ,
\end{equation}
for all $x$-values. The probability density function for the
interaction of a particle after traveling a distance $x$ in the
medium is given by\,\cite{Tavernier:2010}
\begin{equation}
w(x) = \frac{1}{\lambda} e^{-x/\lambda} \ .
\end{equation}
If the incident beam energy per nucleon during hard collision $E(x)$
is above the threshold energy of $\approx 290$ MeV pions are
produced. The pion production multiplicity, $\mu(E) =
\sigma_{\pi}/\sigma_{\rm tot}$, depends greatly on the kinetic
energy of the incident particles as given by Eqn. (\ref{EqnEx2}).
Here $\sigma_{\pi}$ is the pion production cross section, whereas
$\sigma_{\rm tot}$ is total reaction cross section. We discuss
$\mu(E)$ in Sec. \ref{Results}. Then the total number of pions
produced per infalling particle can be calculated as
\begin{equation}
N_{\pi^+}  = \int_{0}^{x_{\rm max}} \mu_{\pi^+}(E) w(x) dx \ ,
\end{equation}
where $x_{\rm max}$ is the range, or the maximum possible distance
the incoming charged particle can penetrate the matter before losing
all of its kinetic energy through electromagnetic energy loss.

\subsection{Transport Optical Depth and Deep Crustal Heating}
\label{EnDeposit}

We want now to demonstrate a first-order rough estimation of the
location in the crust where neutrinos deposit their energies. The
neutrinos moving into the neutron star crust ``forget" their
original direction of motion after a succession of collisions and
having been carried a distance corresponding to their transport mean
free path which can be determined by the neutrino transport optical
depth
\begin{equation}
\label{OptDepth} \tau^{\rm tr} = \int_0^{l} (\sigma_{\nu i}^{\rm tr}
\rho_i + \sigma_{\nu n}^{\rm tr} \rho_n) dl' \ ,
\end{equation}
where $\sigma_{\nu i}^{\rm tr}$ ($\sigma_{\nu n}^{\rm tr}$) is the
neutrino-ion (neutrino-free neutron) transport cross section,
$n_{\mathrm{ion}} = n/A$ is the ion number density in the crust, $A$ is the
number of nucleons in the unit cell, $n_n = N_{\rm f} n_{\mathrm{ion}} $ is
the number density of free neutrons, and $N_{\rm f}$ is the number
of free neutrons in a unit cell. The transport cross section is
defined as
\begin{equation}
\sigma^{\rm tr} = \int d\Omega \frac{d\sigma}{d\Omega}(1-\cos\theta)
\ ,
\end{equation}
with the free-space differential cross section for neutrino-nucleon
elastic scattering given by~\cite{Horowitz:2001xf}
\begin{equation}
\frac{d\sigma_{\nu n}} {d\Omega} =
\frac{G_F^2E_{\nu}^2}{4\pi^2}\left(C^2_{\rm
v}(1+\cos\theta)+C^2_{\rm a}(3-\cos\theta)\right) \ ,
\end{equation}
where $\theta$ is the scattering angle, $E_{\nu}$ is the incoming
neutrino energy, $C_{\rm v}$ is the vector coupling constant, and
$C_{\rm a}$ is the axial vector coupling constant. The neutrino-ion
elastic scattering differential cross section is
\begin{equation}
\frac{d\sigma_{\nu i}} {d\Omega} =
\frac{G_F^2E_{\nu}^2}{4\pi^2}\left(1+\cos\theta\right)\frac{Q^2_{\rm
w}}{4}F(Q^2)^2 \ ,
\end{equation}
where $Q_{\rm w} = N Q^{\rm n}_{\rm w} + Z Q^{\rm p}_{\rm w}$ is the
total weak charge of the ion with  $Q^{\rm n}_{\rm w} =-0.9878$ and
$Q^{\rm p}_{\rm w}=0.0721$, $F(Q^2)$ is the ground state elastic
form factor of the ions\,\cite{Horowitz:2003cz}
\begin{equation}
F(Q^2) = \frac{1}{Q_{\rm w}} \int d^3 r
\frac{\sin{(Qr)}}{Qr}\left[\rho_{\rm n}(r) - \left(1-4
\sin^2{\theta_{\rm W}}\right)\rho_{\rm p}(r) \right] \ ,
\end{equation}
with $Q^2 = 2 E_{\nu}^2 (1-\cos\theta)$ being the four momentum
transfer squared and $\theta_{\rm W}$ is the Weinberg angle.

Notice that both $\sigma_{\nu i}^{\rm tr}$ and $\sigma_{\nu n}^{\rm
tr}$ are functions of the neutrino energy. The neutrino energy
spectrum from stopped pions is well known~\cite{Scholberg:2005qs}.
To determine the neutrino-ion and neutrino-free neutron elastic
scattering cross sections we use the root-mean-square neutrino
energies calculated as
\begin{equation}
E_{\nu}^{\rm rms} =\left(\frac{\int E^2\Phi(E)
dE}{\int\Phi(E)dE}\right)^{1/2} \ ,
\end{equation}
where $\Phi(E)$ is the neutrino flux with energy
$E$~\cite{Scholberg:2005qs}. In particular, we use the
root-mean-squared values of $E_{\nu_{e}}^{\rm rms} =33.3\,
\mathrm{MeV}$ and $E_{\bar{\nu}_{\mu}}^{\rm rms} =37.7\,
\mathrm{MeV}$ for electron and muon neutrinos, respectively (see
Eqn.~(\ref{TotQnu})) and $E_{\nu_\mu}=29.8\, \mathrm{MeV}$.

By definition, $\tau^{\rm tr}$ which is given by Eqn.
(\ref{OptDepth}) represents the number of transport mean free paths
for the neutrino traveling from the surface of the star at $l = 0$
to some inner depth $l$. Neutrinos are assumed to ``forget" their
original direction of motion at a depth of $l = \Delta R$
corresponding to $\tau^{\rm tr} \approx 1$. As a first-order rough
estimate, one can assume neutrinos eventually deposit their energies
around this optical depth.

Electron neutrinos are most likely absorbed in the crust via
inelastic neutrino charged current interactions (\textit{e.g.}
$\nu_e  + n \rightarrow p + e^-$, or $\nu_e  + X^{A}_{Z} \rightarrow
X^{\prime \, A}_{Z+1} + e^-$). Whereas muon (anti)neutrinos deliver
most of their energies through muon (anti)neutrino-electron
scatterings, \textit{i.e.}
\begin{eqnarray*}
&& \nu_{\mu} + e^{-} \rightarrow \nu_{\mu}^{\prime} + e^{-\,\prime}
\ , \\ &&  \bar{\nu}_{\mu} + e^{-} \rightarrow
\bar{\nu}_{\mu}^{\prime} + e^{-\prime} \ .
\end{eqnarray*}
One can estimate the average fractional energy loss 
for muon neutrinos
interacting with electrons at rest as well as for the degenerate gas
of electrons~\cite{Bahcall:1964zzb}.
These estimations show that in about tens of electron scatterings,
the muon (anti)neutrinos will lose most of their energies, and will
eventually leave the star as their energy becomes low enough
corresponding to a very small cross-section, hence a very large
neutrino mean free path.

Another possible channel for neutrinos to deposit their energies
could be through a charged-current reaction $\nu_{\mu} + e^{-}
\rightarrow \mu^{-} + \bar{\nu}_e$. This is because the transport
optical depth corresponds to the density regions in the crust where
muon threshold can be reached. Notice that for non-energetic
neutrinos the muon threshold is usually at a much higher densities.
Other possible channels such as the neutrino-neutron scattering
might not be as important as neutrino-electron scatterings. That is
because although the cross-section for neutrino-neutron scattering
is much bigger than the neutrino-electron scattering, the amount of
energy deposited by a single neutrino-neutron scattering is much
smaller than the one by a single neutrino-electron scattering.
Finally, one can think of inelastic neutral current neutrino-nucleus
scattering as a potentially interesting channel for neutrino energy
deposition. This has been discussed at supernovae neutrino energies
and is believed to be important for supernovae
simulations\,\cite{Langanke:2004vx}. However, more works are needed
to be done to determine whether this is a robust channel for
neutrino energy deposition in the neutron star crust. Thus, it is
safe to conclude that as a first-order rough estimate most of the
energy of $Q_{\nu}$ given by the Eqn.~(\ref{TotQnu}) is delivered to
the inner crust of depths of about $\sim\Delta R$.


\section{Results}
\label{Results}

\subsection{Equations of State of Neutron-Star Matter}
As noted earlier, the multiplicty of pion productions strongly
depends on the kinetic energy of the incoming particles, which in
turn depends on the compactness parameter, $R_{\rm S}/R$. Since the
stellar compactness is strongly sensitive to neutron-star equation
of state, for realistic considerations we will consider a set of the
equations of state that are consistent with current nuclear
experimental and observational constraints. Moreover, a detailed
knowledge of the equation of state of the crust is important in
calculations of the transport optical depth.

The equation of state adopted in this work is composed of several
parts. Matter in the outer crust of the neutron star is organized
into a Coulomb lattice of neutron-rich nuclei embedded in a
degenerate electron gas. The composition in this region is solely
determined by the masses of neutron-rich nuclei in the region of $26
\leq Z \lesssim 40$ and the pressure support is provided primarily
by the degenerate electrons. For this region we adopt the equation
of state by Haensel, Zdunik and Dobaczewski
(HZD)\,\cite{Haensel:1989}. The inner crust begins at the
neutron-drip density of $\rho_{\rm drip} \approx 4 \times 10^{11} \,
\mathrm{g \ cm^{-3}}$. The EOS for the inner crust at mass densities
$\rho > \rho_{\rm drip}$ is, however, highly uncertain and must be
inferred from theoretical calculations. In addition to a Coulomb
lattice and an electron gas, the inner crust now includes a dilute
vapor of quasi-free neutrons. Moreover, at the bottom layers of the
inner crust, complex and exotic structures with almost equal
energies referred to as ``nuclear pasta" have been predicted to
emerge\,\cite{Ravenhall:1983uh, Hashimoto:1984, Lorenz:1992zz}. For
this region we use the EOS by Negele and
Vautherin\,\cite{Negele:1971vb}. The inner crust ends at a mass
density near $\rho_{\rm t} \approx 1.3 \times 10^{14} \, \mathrm{g \
cm^{-3}}$, beyond which the neutron star matter becomes uniform. For
this uniform liquid core region we assume two equations of state
that cover a wide range of uncertainties that currently exist in the
determination of the equation of state of nuclear matter at normal
and supra nuclear densities:

\begin{itemize}
\item The relativistic mean-field
model by Chen and Piekarewicz\,\cite{Chen:2014sca} (\emph{FSU2}),
whose parameters were calibrated to reproduce the ground-state
properties of finite nuclei and their monopole response, as well as
to account for the maximum neutron star mass observed to
date\,\cite{Demorest:2010bx, Antoniadis:2013pzd, Fonseca:2016tux}.
Due to the lack of stringent isovector constraints, the original
\emph{FSU2} predicts a relatively stiff symmetry energy of $J = 37.6
\pm 1.1$ MeV with density slope of $L = 112.8 \pm 16.1$ MeV. It is
known that by tuning two purely isovector parameters of the RMF
model one can generate a family of model interactions that have
varying degrees of softness in the nuclear symmetry energy without
compromising the success of the model in reproducing ground-state
properties\,\cite{Horowitz:2000xj, Fattoyev:2012ch}. Following this
scheme we tuned the purely isovector parameters of the \emph{FSU2}
model to get $J = 31.1$ MeV and $L = 50.0$ MeV and refer to this
model as the \emph{FSU2} (soft). The maximum neutron-star masses
predicted by these models are $2.07 M_{\odot}$ and $2.03 M_{\odot}$,
respectively.

\item The soft and stiff equations of state that agree with the
lower and upper limits of the EOS band derived from microscopic
calculations of neutron matter are based on nuclear interactions
from chiral effective field theory by Hebeler \textit{et
al.}\,\cite{Hebeler:2010jx} (\emph{HLPS}). Notice that the symmetry
energy parameters in this model are $29.7 < J < 33.5$ MeV and $32.4
< L < 57.0$ MeV. Similarly, the maximum stellar masses predicted by
these models are $2.04 M_{\odot}$ and $2.98 M_{\odot}$,
respectively.
\end{itemize}

\begin{table}[h]
\begin{tabular}{|l||c|c|c|c|}
 \hline
 Model                        & $R_{14}$ (km) & $T_{14}$ (MeV) & $R_{20}$ (km) & $T_{20}$ (MeV) \\
 \hline
 \hline
 \emph{FSU2} (soft)           & $12.89$       & $200.5$        & $12.03$       & $377.2$ \\
 \emph{FSU2} (stiff)          & $14.10$       & $178.0$        & $12.95$       & $334.2$ \\
 \emph{HLPS} (soft)           & $~9.95$       & $289.5$        & $~9.68$       & $565.2$ \\
 \emph{HLPS} (stiff)          & $13.59$       & $186.8$        & $14.14$       & $291.6$ \\
\hline
\end{tabular}
\caption{The radii $R$ of a $1.4-$ and $2.0\, \mathrm{M_{\odot}}$
neutron stars as well as the incoming kinetic energies of a nucleon
(with mass of $m_{\rm N} = 939$ MeV) at the surface of neutron stars
$T$ predicted by the four equations of state discussed in the text.}
\label{Table1}
\end{table}

A recent survey on the mass spectrum of compact objects in X-ray
binaries from 19 sources shows that their masses can be anywhere in
the range of $M = (0.9-2.7)\, M_{\odot}$~\cite{Casares:2017jah}.
Note that stars made with stiff equations of state can accelerate
particles to near the pion-production threshold only for more
massive stars, whereas those with soft equations of state allow
particles to gain kinetic energies significantly larger than the
pion-production threshold even for low-mass neutron stars (see Table
\ref{Table1}).

\subsection{Production of $\pi^{+}$ in $p$-$p$, $p$-$Fe$ and $\alpha$-$Fe$ Collisions}

In this subsection we will discuss the sensitivity of the pion
production to the incoming energy of the particle and to the target
nuclei. We assume that the accreted matter (incoming particles) is
composed of either protons or helium. The surface composition of
neutron stars (target nuclei) however remains an outstanding
problem~\cite{Chang:2010sx}. For accreting neutron stars, the upper
layer is likely composed of lighter elements such as hydrogen or
helium, depending on the composition of accreted material from the
companion star.
Ultra-compact low mass X-ray binaries with orbital periods of tens
of minutes accrete from a hydrogen-deficient companion that can be a
He, C-O, or O-Ne-Mg white dwarf, so that the neutron star surface
composition would consist of heavier elements than H or He,
\textit{e.g.} see Ref.\,\cite{intZand:2005sfk}. For the sake of
simplicity, instead of a range of target nuclei, we assume only two
types of the target nuclei, protons and $Fe$, and only a select
nuclear collisions: $p$-$Fe$, $p$-$p$, and $\alpha$-$Fe$.

Charged pion production from the interaction of proton beams with
some selected nuclei have been measured at incident energies of 585
MeV~\cite{Crawford:1979ia}, 730 MeV~\cite{Cochran:1972pq}, as well
as at 800 and 1600 MeV~\cite{Denes:1983mw, Lemaire:1991xs}.
Inclusive pion production at lower incident energies of 330, 400,
and 500 MeV from proton-nucleus collisions ($^{12}C$ and $^{138}U$)
nuclei have also been measured. However, measurements of pion
production cross sections at medium beam energies for proton-nucleus
collisions are still incomplete.

\begin{figure}[h]
 \begin{center}
\includegraphics[width=1.0\columnwidth,angle=0]{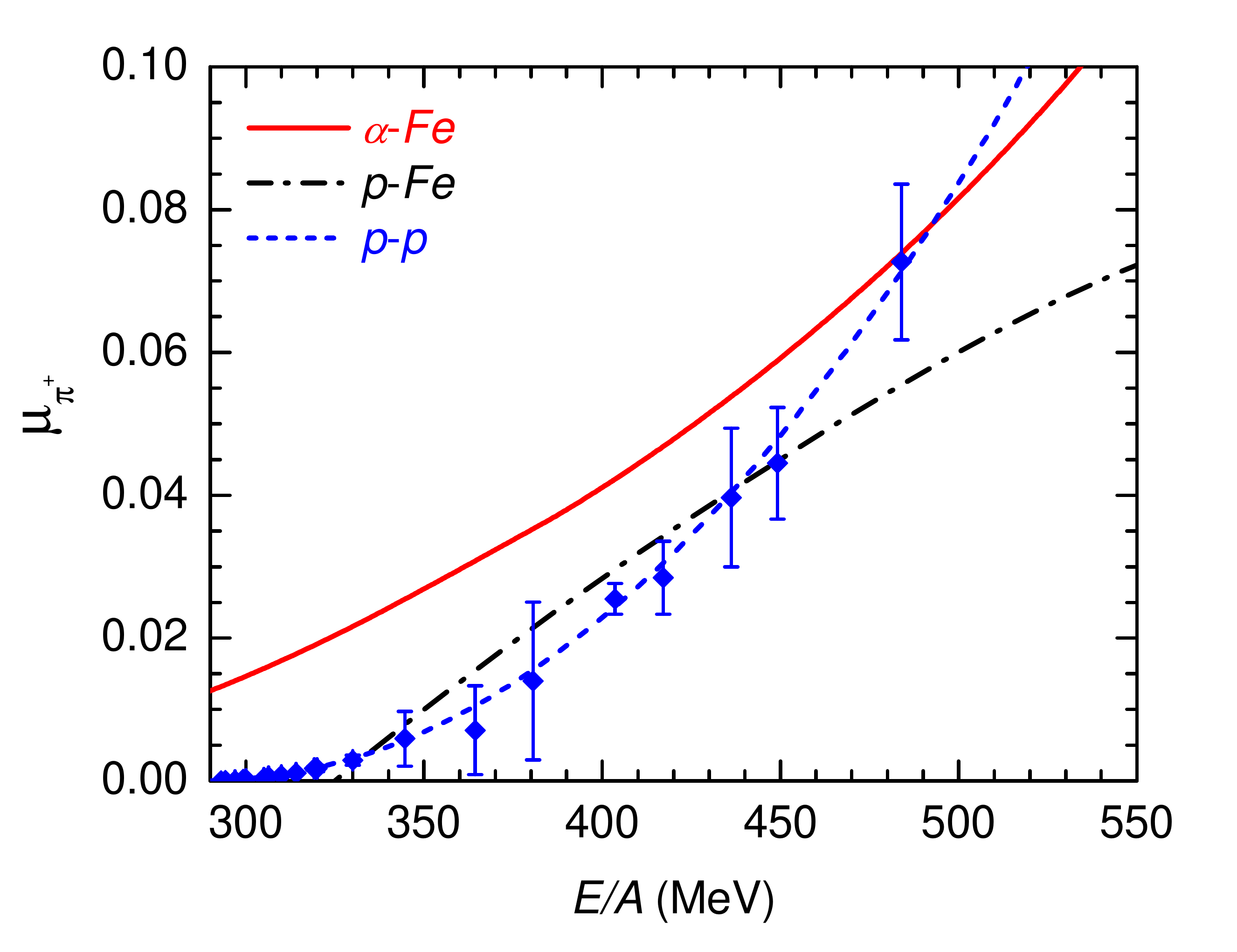}
 \end{center}
  \vspace{-0.4cm}
   \caption{(Color online)
  The multiplicity of pion production as a function of beam energy for
  $p$-$Fe$ collisions (dash-dotted black line) estimated using the Monte-Carlo results of
  Ref.~\cite{Burman:1989dq} and for $p$-$p$ collisions (dashed blue line) which is a polynomial fit to
  the experimental data from~\cite{Daehnick:1995gw, Hardie:1997mg, Flammang:1998bb, Schwaller:1979eu}
  (blue diamond symbols). Also shown is $\pi^{+}$ multiplicities from the
  isospin-dependent Boltzmann-Uehling-Uhlenbeck (IBUU) transport
  simulations in the $\alpha$-$Fe$ collision (solid red line).}
 \label{Fig1}
\end{figure}

\begin{figure}[t]
 \begin{center}
\includegraphics[width=1.0\columnwidth,angle=0]{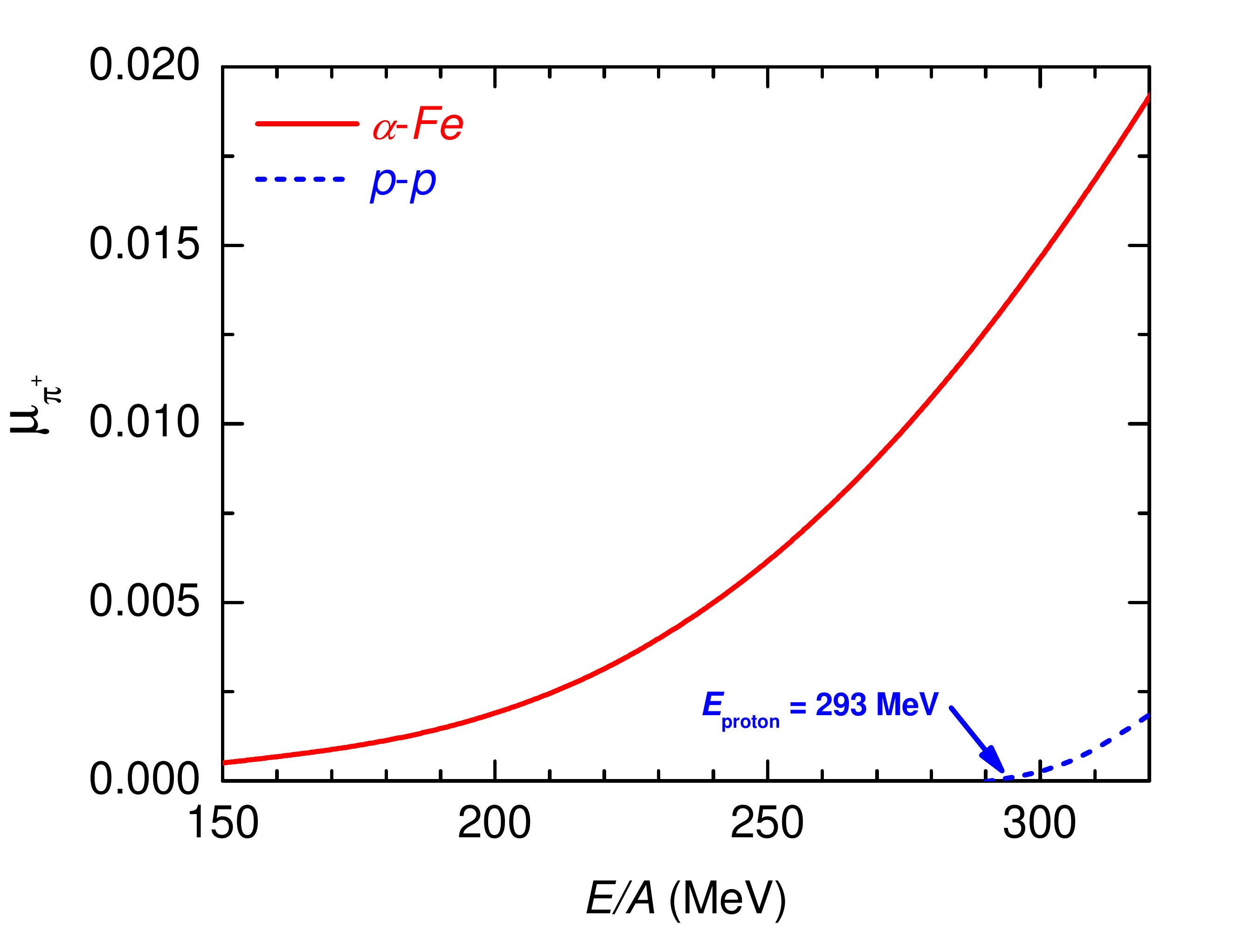}
 \end{center}
  \vspace{-0.4cm}
   \caption{(Color online)
  Same as Fig. 1 but now for `subthreshold' energies.}
 \label{Fig2}
\end{figure}
Based on the available experimental data Ref.~\cite{Burman:1989dq}
performed a Monte-Carlo simulation to evaluate the total pion
production cross sections at various proton beam energies on
selected nuclei. Using these Monte-Carlo data and the $p$-$Fe$
reaction cross sections we estimated pion multiplicities for
$p$-$Fe$ collision at incident beam energies above 325 MeV, see
Fig.~\ref{Fig1}. Note that the pion production cross section in
Monte-Carlo simulations is assumed to go to zero at energies below
325 MeV~\cite{Burman:1989dq}. It is important to mention, however
that pions can be produced at subthreshold energies via the
excitation and decay of $\Delta$-resonances (See
Ref.~\cite{Tsang:2016foy} and references therein). The Fermi motion
of nucleons in nuclei can also greatly enhance the pion production
cross sections in the vicinity of the threshold
energy~\cite{Bertsch:1977ps, Sandel:1979tg}. Despite efforts to
measure subthreshold pion production in the past (See,
Ref.~\cite{Miller:1993zw} and references therein), regrettably,
experimental data on this front still remains incomplete.

\begin{figure}[t]
 \begin{center}
\includegraphics[width=1.0\columnwidth,angle=0]{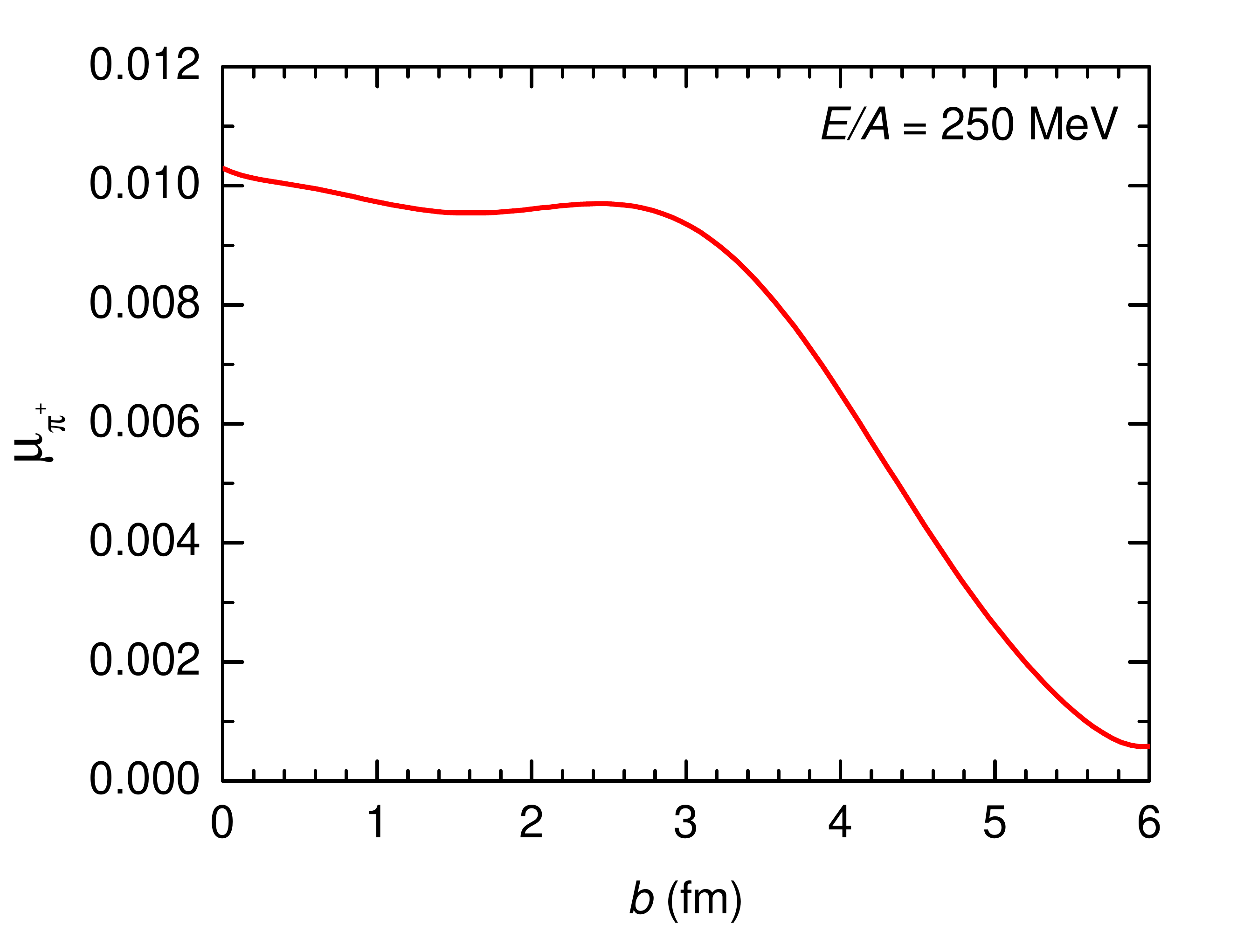}
 \end{center}
  \vspace{-0.4cm}
   \caption{(Color online)
  The $\pi^{+}$ multiplicity in $\alpha$-Fe collisions versus impact
  parameter $b$ for incident $\alpha$ energies per nucleon of $E/A =
  250$ MeV calculated using IBUU transport simulations. }
 \label{Fig3}
\end{figure}

For stars that accrete hydrogen from the companion, the $p$-$p$
collision becomes an interesting
case~\cite{Bildsten:1992ApJ,Bildsten:1993ApJ} to study the pion
production. Fortunately, there are sufficient experimental data
available on the pion production in $p$-$p$ collisions. Using the
experimental data from~\cite{Daehnick:1995gw, Hardie:1997mg,
Flammang:1998bb, Schwaller:1979eu} we plot the multiplicity of pion
production as a function of the proton beam energy from the $p$-$p$
collisions (see Fig.~\ref{Fig1}). The current experimental
error-bars are in the order of 25\% for most of these measurements,
except for few cases when beam energies are in the range of $344.6 <
E < 380.6$ MeV, the relative error-bars are as large as
$\simeq$80\%.

On the other hand, experimental measurements of pion production in
$\alpha$-$Fe$ collisions are still missing. For this we use the IBUU
transport model to calculate $\pi^{+}$ multiplicities at various
incident beam energies per nucleon and impact parameters. For a
detailed description of this transport model we refer the reader to
Ref.~\cite{Li:2002qx, Li:2002yda}. The results are presented in
Fig.~\ref{Fig1} alongside $p$-$Fe$ and $p$-$p$ collisions. In
Fig.~\ref{Fig2} we display predicted $\pi^{+}$ multiplicities as a
function of the incident beam energies per nucleon for subthreshold
energies. For comparison, we also show the results from $p$-$p$
collisions in the lower right corner. In this model all subthreshold
pions are produced from decays of low-mass $\Delta$(1232) resonances
formed in nucleus-nucleus inelastic collisions. While the pion
production cross sections drop sharply when the energy per nucleon
is below threshold, there is an appreciable pion production cross
section at incident beam energies as low as 150 MeV per nucleon
mostly due to Fermi motion of nucleons in $Fe$. Note that pion
production in heavy-ion collisions depends on the EOS and the ratio
of charged pions on the nuclear symmetry energy used. In this
exploration study, we use a momentum-independent potential
corresponding to a stiff EOS with $K_0=380$ MeV and a symmetry
energy that is linear in density. In our calculation for
$\alpha$-$Fe$ collisions, we used $\mu_{\pi^{+}}$ multiplicities
averaged over the impact parameter $b$:
\begin{equation}
\mu_{\pi^{+},\,\rm ave} = \frac{\int_0^{b_{\rm max}}
\mu_{\pi^{+}}(b) b db}{\int_0^{b_{\rm max}} b db} \ ,
\end{equation}
whose dependence is plotted in Fig.~\ref{Fig3}. For head-on
collisions with $b = 0\, \mathrm{fm}$ the pion-production cross
section is obviously much larger. In generating these data we run
$100,000$ events for most cases, except for low incident energies,
where we used up to $300, 000$ events to have a better statistics.
The statistical error-bars for these simulations are of the order of
$\approx 30\%$. It is worth noting that uncertainties coming from
different EOS models and the density dependence of the symmetry
energy are of the same order as the statistical errors quoted above.
The detailed model dependencies near the pion production threshold
are addressed in Ref.~\cite{Li:2015hfa}.

\subsection{Neutrino Energy Deposition in the Inner Crust}
\begin{table}[h]
\begin{tabular}{|l||c|c|c|}
 \hline
 Model                        & $Q_{\nu}^{pp}$ (MeV) & $Q_{\nu}^{p \rm Fe}$ (MeV) & $Q_{\nu}^{\alpha \rm Fe}$ (MeV) \\
 \hline
 \hline
 \emph{FSU2} (stiff)          & $0.009$  [$0.000$]   & $0.005$ [$0.000$]          & $0.258$ [$0.004$] \\
 \emph{FSU2} (soft)           & $0.069$  [$0.000$]   & $0.131$ [$0.000$]          & $0.493$ [$0.009$] \\
 \emph{HLPS} (stiff)          & $0.000$  [$0.000$]   & $0.000$ [$0.000$]          & $0.114$ [$0.006$] \\
 \emph{HLPS} (soft)           & $2.114$  [$0.000$]   & $1.963$ [$0.000$]          & $2.822$ [$0.109$] \\
\hline
\end{tabular}
\caption{The total energy per accreted nucleon deposited by neutrinos
in the inner crust for a 2.0 [1.4]$\, \mathrm{M_{\odot}}$ neutron star using
the four equations of state discussed in the text and for three
possible reactions: $pp$, $p$-Fe, and $\alpha$-Fe.} \label{Table2}
\end{table}
\begin{table*}[t]
\begin{tabular}{|l||c|c|c|c|c|c|c|c|c|c|}
 \hline
 Model                        & $M$           & $\rho_{\bar{\mu}}$      & $y_{\bar{\mu}}$         & $\Delta R_{\bar{\mu}}$  & $\rho_{e}$              & $y_{e}$                 & $\Delta R_{e}$ & $\rho_{\mu}$            & $y_{\mu}$               & $\Delta R_{\mu}$  \\
                              & ($M_{\odot}$) & ($10^{12} \, \mathrm{g \ cm^{-3}}$) & ($10^{16} \, \mathrm{g \ cm^{-2}}$) & (km)                    & ($10^{12} \, \mathrm{g \ cm^{-3}}$) & ($10^{16} \, \mathrm{g \ cm^{-2}}$) & (km)           & ($10^{12}\, \mathrm{g \ cm^{-3}}$) & ($10^{16} \, \mathrm{g \ cm^{-2}}$) & (km) \\
 \hline
 \hline
 \emph{FSU2} (soft)           & $1.4$         & $~5.0$                 & $3.2$                 & $0.60$                 & $~6.1$                 &  $3.9$                & $0.62$        & $~7.1$                 & $4.7$                 & $0.63$ \\
                              & $2.0$         & $~9.7$                 & $4.0$                 & $0.31$                 & $11.3$                 &  $5.0$                & $0.31$        & $12.9$                 & $6.1$                 & $0.32$ \\
 \emph{HLPS} (soft)           & $1.4$         & $~7.4$                 & $4.0$                 & $0.34$                 & $~8.5$                 &  $4.6$                & $0.35$        & $10.4$                 & $5.6$                 & $0.35$ \\
                              & $2.0$         & $16.5$                 & $5.0$                 & $0.17$                 & $18.7$                 &  $6.1$                & $0.17$        & $20.7$                 & $7.4$                 & $0.18$ \\
\hline
\end{tabular}
\caption{The location in the neutron star inner crust, where
neutrino transport optical depth equals $\tau^{\rm tr} = 1$. Here
$\rho_{\bar{\mu}}$, $\rho_{e}$, and $\rho_{\mu}$ are the mass
density of the neutron-star matter in the crust where neutrinos with
different flavors are first scattered (or absorbed), $y = P/g$ and
$\Delta R$ are the corresponding column depth~\cite{Brown:2009kw}
and radial depth, where $P$ is the local pressure and $g$ is the
local gravitational acceleration~\cite{Thorne:1977} at $r = R-\Delta
R$.} \label{Table3}
\end{table*}
\begin{figure}[h]
 \begin{center}
\includegraphics[width=1.0\columnwidth,angle=0]{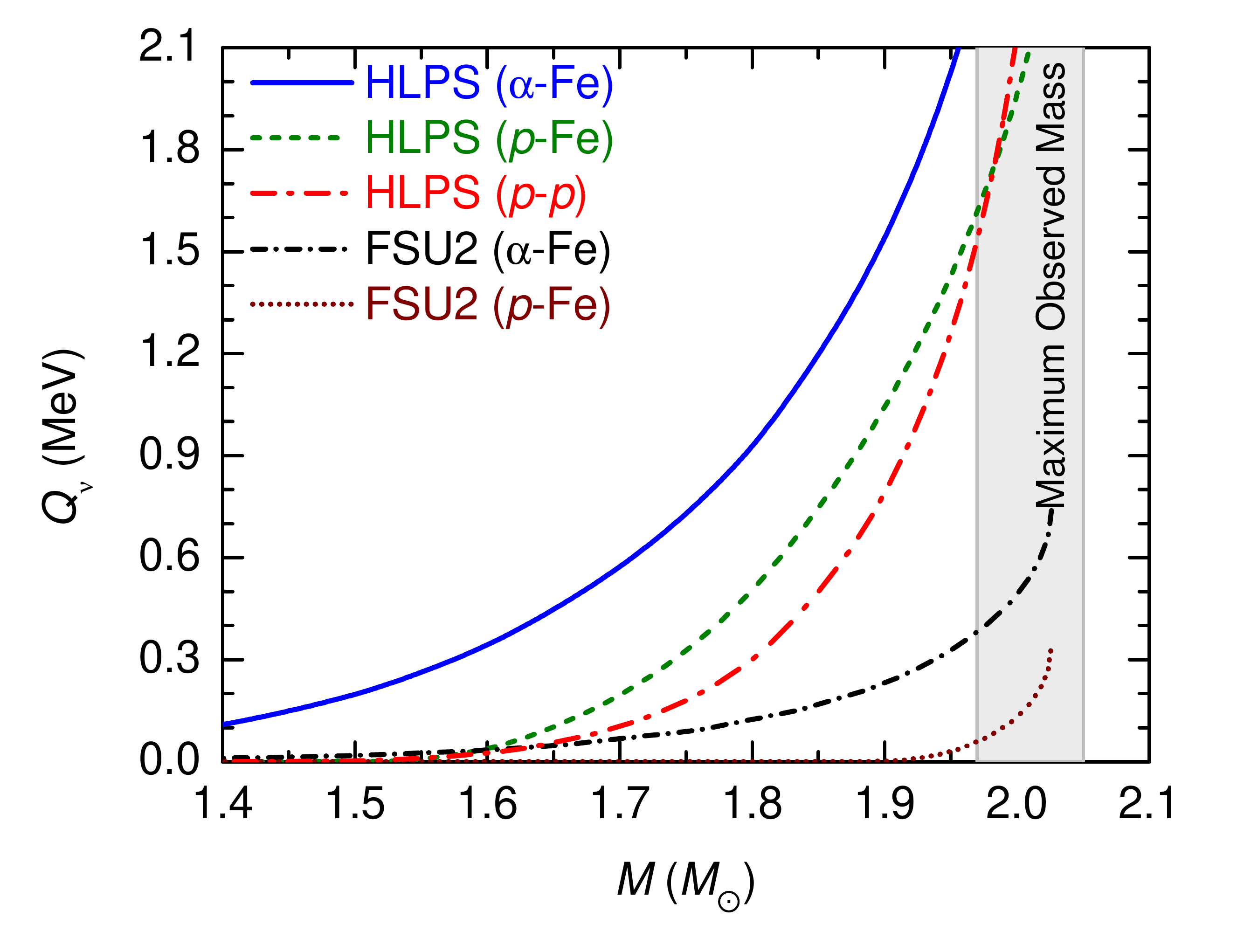}
 \end{center}
  \vspace{-0.4cm}
   \caption{(Color online)
  The total energy deposited by neutrinos into the inner crust as a function of neutron start mass for
  the soft EOSs discussed in the text: \emph{HLPS} (soft) and \emph{FSU2} (soft).}
 \label{Fig4}
\end{figure}
We now calculate the total energy carried by neutrinos into the
inner crust. In Table~\ref{Table2} we present results for a $2 \,
\mathrm{M_{\odot}}$ and for a $1.4\, \mathrm{M_{\odot}}$ (in square
brackets) neutron star and the four equations of state discussed in
the text. In Fig.~\ref{Fig4} we display the full results as a
function of the neutron star mass for soft equations of state only.
As is also evident from Table~\ref{Table1}, low-mass neutron stars
can accelerate the infalling matter to energies of about the
pion-production threshold only. Therefore the result is highly
sensitive to the pion production cross section around threshold
energies. This result calls for improved experimental measurements
of pion production in proton-proton collisions, as well as for pion
production in $p$-$Fe$ and $\alpha$-$Fe$ collisions for beam
energies per nucleon in the range of $150$ to $600$ MeV.

\begin{figure}[h]
 \begin{center}
\includegraphics[width=1.0\columnwidth,angle=0]{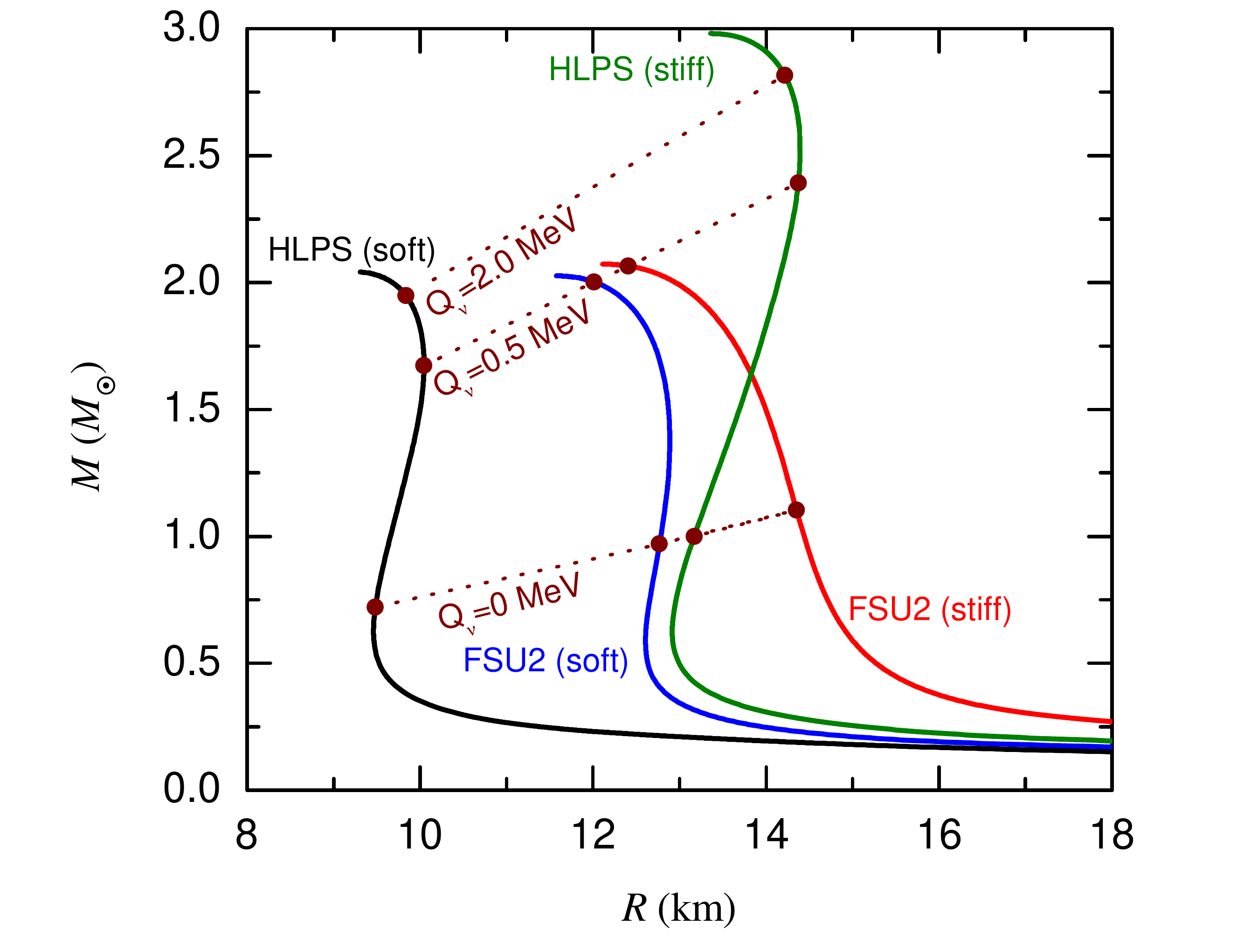}
 \end{center}
  \vspace{-0.4cm}
   \caption{(Color online). Mass-vs-Radius relation predicted by the four EOS
   models discussed in the text. The contours of constant
   neutrino heat deposits are shown with dotted curves.}
 \label{Fig5}
\end{figure}
Moreover, there is a strong sensitivity of the pion production to
the equation of state employed in determination of stellar
structure. In particular, if the equation of state is very
stiff---such as the \emph{HLPS} (stiff)---then even for a $2\,
\mathrm{M_{\odot}}$ neutron star the incoming particles are not
accelerated enough to produce pions (See Tables~\ref{Table1} and
~\ref{Table2}). On the other hand, if the equation of state is soft,
then for a $M = 2 M_{\odot}$ neutron star the energy deposited by
neutrinos can be as large as $Q_{\nu} \simeq 2.8$ MeV per accreted
nucleon. The result is more pronounced if helium is being accreted
onto the surface of neutron star, mainly because the IBUU
simulations suggest that a substantial amount of pions can be
produced at subthreshold beam energies. In Fig.~\ref{Fig5} we plot
the Mass-vs-Radius relation predicted by the four equations of
state, where the contours of $Q_{\nu}=$ $0$, $0.05$, and $2.0$ MeV's
coming from $\alpha$-$Fe$ collisions are plotted. As evident from
the figure, the heating gets more pronounced for massive and/or
compact stars only.

\begin{figure}[h]
 \begin{center}
\includegraphics[width=1.0\columnwidth,angle=0]{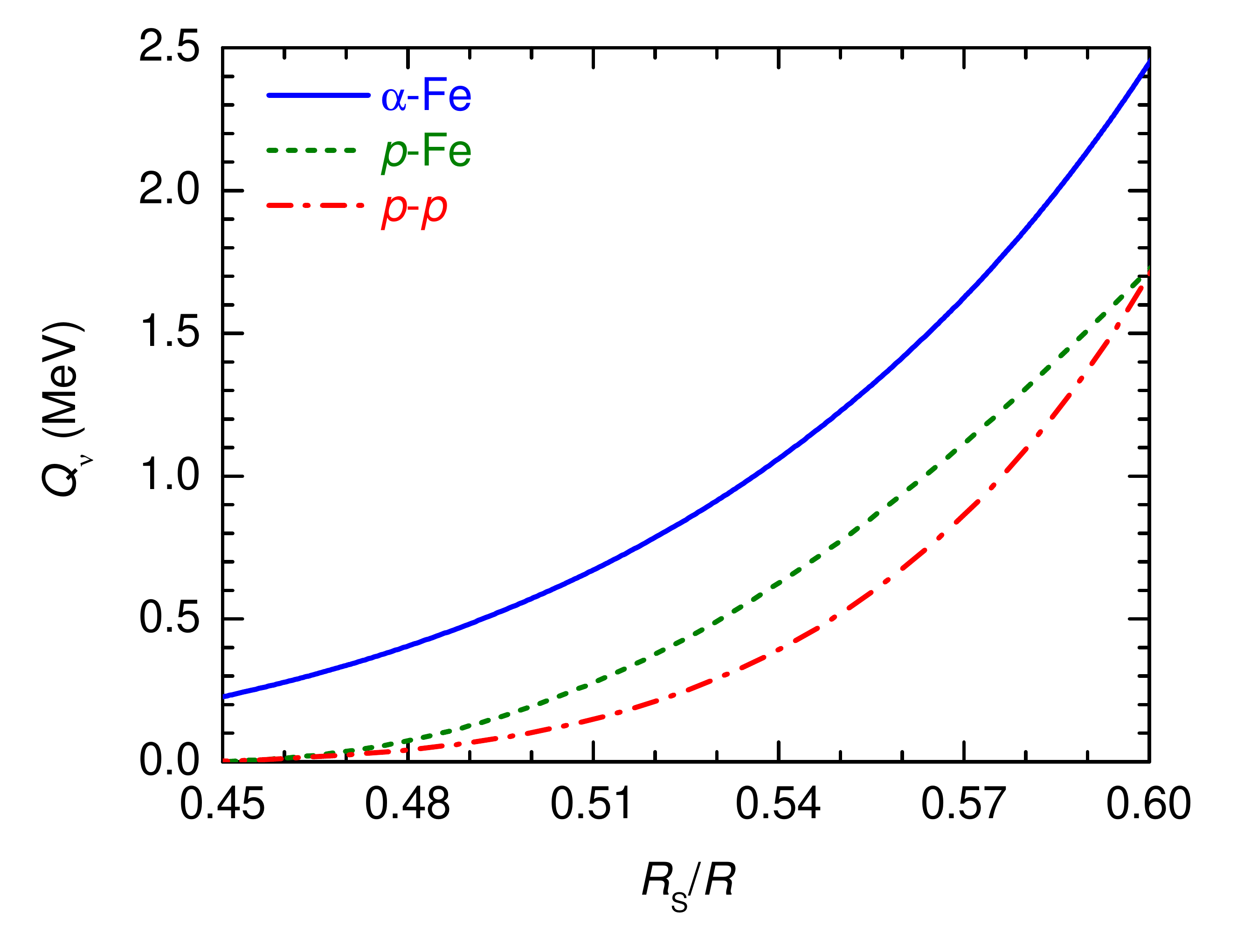}
 \end{center}
  \vspace{-0.4cm}
   \caption{(Color online)
  The total energy deposited by neutrinos into the inner crust as a function of neutron star compactness, $R_{\rm S}/R$.}
 \label{Fig6}
\end{figure}
Finally, let us investigate the impact of the neutron star's
compactness parameter on the amount of heat deposition. While stars
built with the \emph{HLPS} (stiff) equation of state may not
accelerate the infalling matter to high kinetic energies for
low-mass stars, it can certainly do so for very massive neutron
stars. In particular, for a $M = 2.8 M_{\odot}$, the total energy
deposit is $Q_{\nu} = 1.89$ $(1.33)$ MeV, when $\alpha$-Fe ($p$-Fe)
collisions take place at the surface (Also see Fig.\,\ref{Fig5}). To
cover all possible equations of state in Fig.~\ref{Fig6} we plot the
results as a function of the compactness parameter.


We find that the heat deposition from neutrinos is comparable with
other previously known sources of deep crustal heating such as from
pycnonuclear fusion reactions. In Table \ref{Table3} we present our
results for the approximate location in the neutron star inner
crust, where neutrino transport optical depth equals to 1. The
results are presented for the more interesting cases of soft
equations of state only, where $Q_{\nu}$ is significant even for
moderate-mass neutron stars. Depending on the mass of the star and
the EOS model used, the energy of $Q_{\nu}$ is delivered to the
regions of the inner crust where mass densities are of the order
$10^{12}$ to $10^{13} \, \mathrm{g \ cm^{-3}}$. For example, for a
$2 \, \mathrm{M_{\odot}}$ neutron star this would correspond to mass
densities of $9.69 < \rho_{12} < 20.74$ in units of $10^{12} \,
\mathrm{g \ cm^{-3}}$, or equivalently to baryon densities of $0.036
< \rho/\rho_0 < 0.078$, where $\rho_0 \approx 2.66 \cdot 10^{14}$ g
cm$^{-3}$ is the nuclear saturation density.

\begin{figure}[h]
\begin{center}
\includegraphics[width=1.0\columnwidth,angle=0]{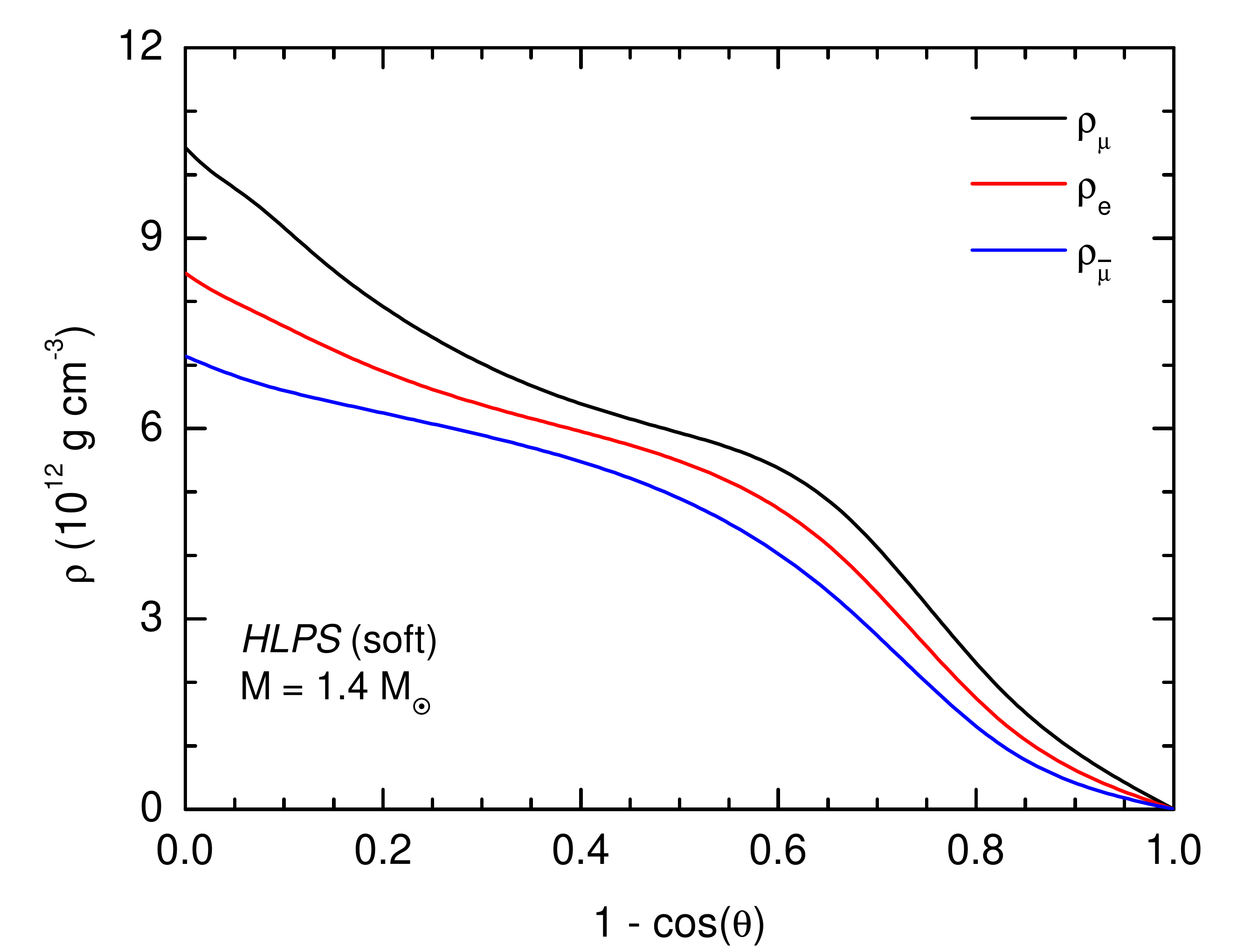}
\end{center}
 \vspace{-0.4cm}
\caption{(Color online) The location in the neutron star inner
crust, where neutrino transport optical depth equals $\tau^{\rm tr}
= 1$ for different angles of incidence $\theta$.}
 \label{Fig7}
\end{figure}

Obviously, $\tau^{\rm tr} = 1$ is only a rough estimate for the
location of the neutrino heat deposition. Moreover, Table
\ref{Table3} assumes that half of the neutrinos produced from the
decay of stopped pions travel radially inward. In reality, the decay
is isotropic and therefore it is worth to analyze the approximate
location of heat delivery as an angle of incidence of neutrinos. The
fraction of the number of neutrinos within a cone with apex angle
$2\theta$ to the total number of neutrinos is equal to $x =
\frac{1}{2} (1-\cos\theta)$. Here $\theta = 0$ corresponds to the
angle of incidence in the radial direction, whereas $\theta =
\frac{\pi}{2}$ corresponds to the direction horizontal to the
surface. In Fig.~\ref{Fig7} we display the location of heat
deposition as a function of $1-\cos\theta$ for a $1.4 M_{\odot}$
neutron star using \emph{HLPS} (soft) EOS. The result shows that
most of neutrinos are delivered to the deep region of the crust, and
only a small fraction of them scatter at shallower regions.

Note that in our calculations above we did not take into account
additional redshift effects as neutrinos go deeper into the crust.
The effective neutrino energy should slightly increase due to the
gravitational redshift by a factor of
$e^{-\phi(r)/c^2}\sqrt{1-R_{\rm S}/R}$, where $\phi(r)$ is the local
gravitational potential~\cite{Thorne:1977}. However our calculations
show this effect is $\lesssim 2\%$ because the crust is thin.

\subsection{Observational Implications}

In this section, we will discuss observational implications of the
extra heating from neutrinos in the presence of other sources of
deep crustal heating. We will first discuss general implications for
neutron stars cooling observation and then apply our result to a
particular neutron star MXB~1659-29.

\subsubsection{Cooling neutron stars}

Non-equilibrium nuclear reactions during active accretion heat the
neutron star crust out of thermal equilibrium with the core. When
accretion stops, the crust cools toward thermal equilibrium with the
core~\cite{Brown:1998qv, Ushomirsky:2001pd, Rutledge:2002,
Cackett:2008tq}. Crust cooling is observed as a quiescent X-ray
light curve, with one of the most well studied examples being the
cooling transient MXB 1659-29~\cite{Brown:1998qv, Ushomirsky:2001pd,
Cackett:2008tq, Brown:2009kw}. Cooling observations at successively
later times into quiescence probes successively deeper layers in the
crust with increasingly longer thermal times~\cite{Brown:2009kw}. In
particular, it was shown that about a year into quiescence the shape
of the cooling light curve is sensitive to the physics at mass
densities greater than neutron drip $\rho > \rho_{\rm drip}$
corresponding to the inner crust~\cite{Page:2012zt}. This suggests
that cooling light curves of neutron stars in low-mass X-ray
binaries one-to-three years after accretion outbursts should be
sensitive to the additional deep crustal heating by
neutrinos~\cite{Brown:2009kw}.

Comparing our results with the heat released from pycnonuclear
fusion reactions~\cite{Haensel:2007wy} we notice that not only are
they of the same order, but also the heat is deposited in the same
density regions (crust layer). Subsequently we calculated the column
depths where neutrinos are first scattered, $y = \int_r^{\infty}
\rho dr \approx P/g$~\cite{Brown:2009kw}. Here $P$ is the local
pressure  and the neutron star's surface gravity is $g =
(GM/R^2)(1-2GM/Rc^2)^{-1/2}$.
We find that the column depth values lie in the range of $4.0 \cdot
10^{16} < P/g < 7.4 \cdot 10^{16} \, \mathrm{g \ cm^{-2}}$ (See
Table \ref{Table3}). Since the amount of heat deposited for massive
stars is comparable to the heat released from pycnonuclear
reactions, the observation of cooling light curves, in particular,
could be used to help distinguish massive stars from the low-mass
stars.

\begin{figure}[h]
 \begin{center}
\includegraphics[width=1.0\columnwidth,angle=0]{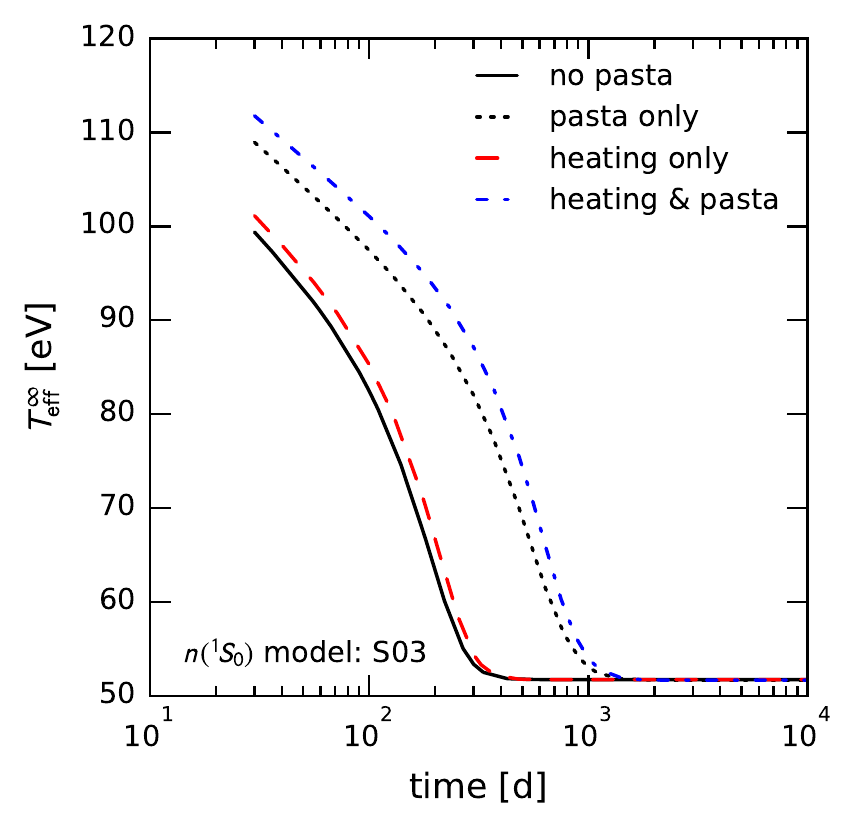}
 \end{center}
  \vspace{-0.4cm}
   \caption{(Color online)
  The redshifted surface temperature in units of $eV$ as a function of time into quiescence. For the details
  of each curve please refer to the text.}
 \label{Fig8}
\end{figure}
To analyze the sensitivity of crustal heating by neutrinos on the
cooling curves, we simulate the thermal evolution of a $2 \,
\mathrm{M_{\odot}}$ neutron star crust using the thermal evolution
code \verb"dStar"~\cite{Brown:2015Code}, which solves the fully
general relativistic heat diffusion equations
\begin{eqnarray}
\frac{\partial}{\partial t} \left(T e^{\phi/c^2}\right) &=&
e^{2\phi/c^2}  \frac{\epsilon_{\rm in} - \epsilon_{\rm out}}{C} - \frac{\frac{\partial}{\partial r} \left(L e^{2\phi/c^2} \right)}{4 \pi r^2 \rho C (1+z)}  \ , \\
L e^{2\phi/c^2}  &=& -\frac{4 \pi r^2 K e^{\phi/c^2}}{1+z}
\frac{\partial}{\partial r} \left(T e^{\phi/c^2}\right) \ ,
\end{eqnarray}
where $\epsilon_{\rm in}$ is the nuclear and/or neutrino heating
emissivity and $\epsilon_{\rm out}$ is the neutrino emissivity from
the core, $C$ is the specific heat, $K$ is the thermal conductivity
and $1+z = [1-2GM/(rc^2)]^{-1/2}$ is the gravitational redshift
factor. The detailed microphysics of the crust is discussed in
Ref.~\cite{Brown:2009kw} and the parameters of the cooling model are
described in Ref.~\cite{Deibel:2016vbc}. In particular, the core
neutrino emissivity $\epsilon_{\rm out}$ includes the modified and
direct Urca reactions that may impact quiescent crust cooling at
late times depending on the core's heat
capacity~\cite{Brown:2017gxd}. Though the core's heat capacity
remains unknown, long term monitoring observations can be used to
place a lower limit on its value~\cite{Cumming:2016weq}. This model
also assumes an impurity parameter of $Q_{\rm imp} = 1.0$ throughout
the crust, which is defined as
\begin{equation}
Q_{\rm imp} = \frac{1}{n_{\rm ion}} \sum_{i} n_{i} \left(Z_i -
\langle Z \rangle\right)^2 \ ,
\end{equation}
where $n_i$ is the
number density of the nuclear species with $Z_i$ number of protons,
and $\langle Z \rangle$ is the average proton number of the crust
composition.

In Fig.~\ref{Fig8} we display the crust cooling curves for four
possible cases. The solid black curve corresponds to the case
without heat deposition from neutrinos in the inner crust with
$Q_{\rm imp} = 1.0$. The red dashed curve corresponds to the case
when a $2.0$ MeV per accreted nucleon heat source is deposited at
density regions of $10^{12} <\rho < 10^{13} \, \mathrm{g \
cm^{-3}}$. The crust temperature is marginally increased by the
neutrino heating because most of the additional heat is transported
into the core.

We then examine two cases, with and without neutrino heating, but
including a nuclear pasta layer in the inner crust. It is expected
that nuclear pasta forms at densities above $\rho > 8.0 \cdot
10^{13} \, \mathrm{g \ cm^{-3}}$ corresponding to the bottom layers
of the inner crust. The thermal conductivity of nuclear pasta could
be small, corresponding to a large impurity
parameter~\cite{Horowitz:2014xca}. The black short-dashed curve
shows the case of no neutrino heating, but $Q_{\rm imp} = 20$ at
densities of $\rho > 8.0 \cdot 10^{13} \, \mathrm{g \ cm^{-3}}$
corresponding to nuclear pasta. Finally, in blue dash-dotted line we
display a cooling curve that includes both nuclear pasta and the
heat depositiion from neutrinos. The crust temperature is higher in
these two cases, because the low thermal conductivity of the nuclear
pasta layer prevents a large portion of heat from diffusing into the
core.

As evident from Fig.~\ref{Fig8}, the additional heat source can make
a noticeable change in the cooling light curves. The cooling rate
depends on many other factors and in particular strongly depends on
the crust thickness, which is usually small for massive stars.
Moreover, as illustrated in Fig.~\ref{Fig8} the low thermal
conductivity corresponding to the nuclear pasta can strongly affect
thermal diffusion time maintaining a temperature gradient between
the neutron star's inner crust and core for several hundred days
into quiescence~\cite{Deibel:2016vbc}. Note that all of the above
models use the same pairing gap model of neutron superfluid in the
$^1S_0$ singlet state with the critical temperature profile given by
Schwenk \textsl{et al.}~\cite{Schwenk:2002fq}. We have also tested
other superfluid pairing gap models, such as the one by Gandolfi
\textsl{et al.}~\cite{Gandolfi:2008id}, and the results are
qualitatively similar to Fig.~\ref{Fig8}.

\begin{figure}[h]
 \begin{center}
\includegraphics[width=1.0\columnwidth,angle=0]{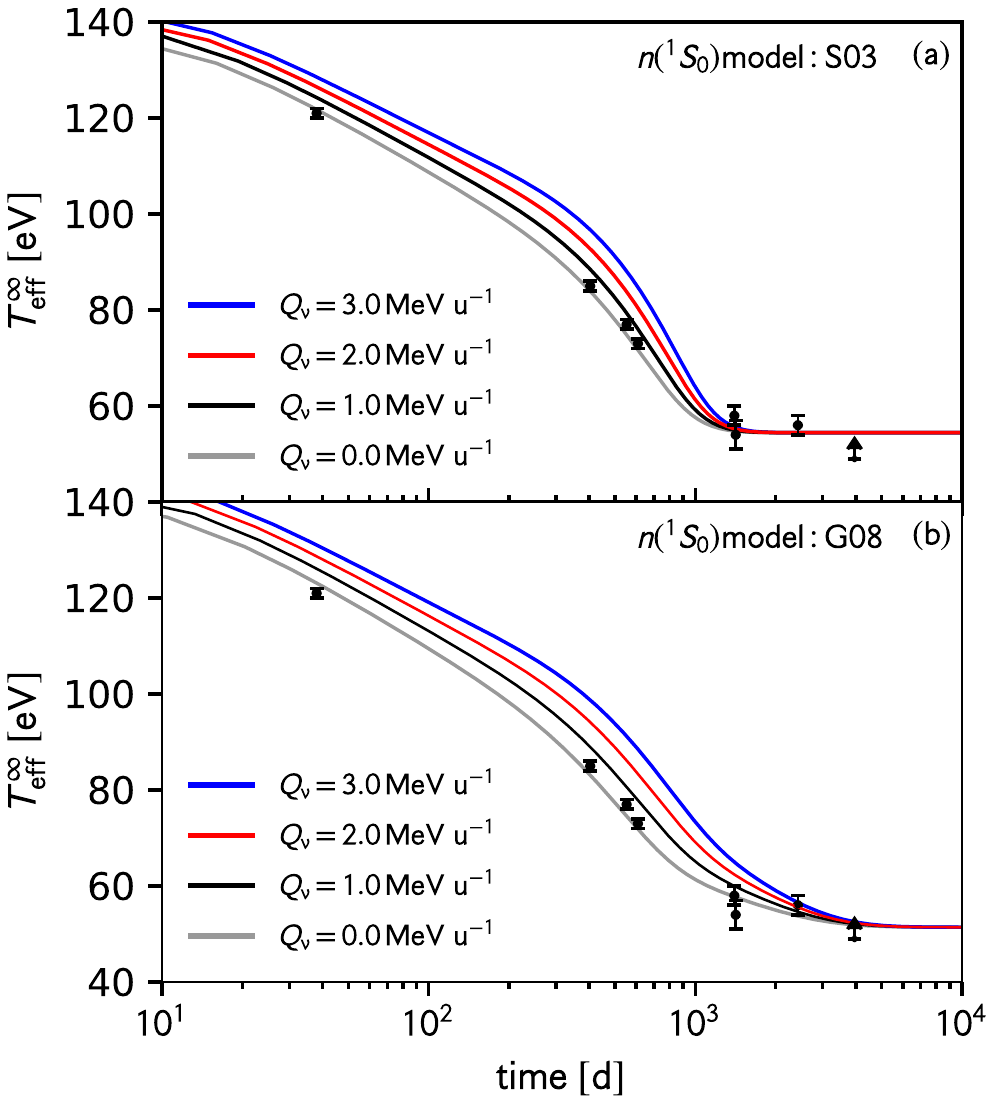}
 \end{center}
 \vspace{-0.4cm}
   \caption{(Color online)
  Observed cooling light curve of MXB~1659-29 (black data points)\,\cite{Cackett:2013wva} and thermal evolution models for three cases of neutrino heating.
  (a) Crust cooling models with the S03 pairing gap and no nuclear pasta. The models use an impurity parameter of $Q_{\rm imp} =3$ for the entire crust
  and a core temperature of $T_{\rm core} = 4.2 \times 10^7 \, \mathrm{K}$.
  (b) Crust cooling models with the G08 pairing gap and nuclear pasta. The models use an impurity parameter of $Q_{\rm imp} =1$ for the crust and
   a $Q_{\rm imp} = 20$ layer representing nuclear pasta at mass densities $\rho \geq 8 \times 10^{13} \, \mathrm{g \ cm^{-3}}$. These models have
   a core temperature of $T_{\rm core} = 3.75 \times 10^7 \, \mathrm{K}$.
   }
 \label{Fig9}
\end{figure}

\subsubsection{Crust cooling in MXB~1659-29}

As described above, neutrino deep crustal heating will noticeably
increase the crust temperature and the shape of the cooling light
curve. Here we investigate the impact of extra heating from
neutrinos on the particular case of MXB~1659-29 that entered
quiescence after an $\approx 2.5 \, \mathrm{year}$ accretion
outburst~\cite{Cackett:2013wva,wijnands2003,wijnands2004} and cooled
for $\approx 4000 \, \mathrm{days}$ before entering outburst once
more~\cite{negoro2015}. The late time cooling observations probe the
thermal properties of the inner crust and make MXB~1659-29 an
interesting test case for neutrino heating.

Our thermal evolution model of MXB~1659-29 uses a $M = 1.6 \,
\mathrm{M_{\odot}}$ and $R = 11.2 \, \mathrm{km}$ neutron star at
the observed outburst accretion rate of $\dot{m} \approx 0.1 \,
\dot{m}_{\rm Edd}$. The model includes a $Q=1\, \mathrm{MeV}$ per
accreted nucleon shallow heat source between $y=2 \times 10^{13} \,
\mathrm{g \ cm^{-2}}$ and $y = 2 \times 10^{14} \, \mathrm{g \
cm^{-2}}$, consistent with the findings from \cite{Brown:2009kw}.
Using a model without nuclear pasta, the cooling light curve is fit
with an impurity parameter for the entire crust of $Q_{\rm imp} = 3$
and the S03 pairing gap~\cite{Schwenk:2002fq}. We then test
representative values of neutrino heating: $1, 2,$ and $3\,
\mathrm{MeV}$ per accreted nucleon. As can be seen in
Fig.~\ref{Fig9}, the model fit with $Q_{\rm imp} = 3$ becomes
inconsistent with the observational data once neutrino heating is
added to the inner crust. In order to reestablish a fit, the crust
impurity parameter must be lowered to $Q_{\rm imp} < 3$
(corresponding to a higher crust thermal conductivity) as $Q_{\nu}$
increases.

Alternatively, the cooling of MXB~1659-29 may be fit with a nuclear
pasta layer in the crust if the G08 pairing gap model is
used~\cite{Gandolfi:2008id}, as we demonstrate in panel (b) of
Fig.~\ref{Fig9}. In this case, the low thermal conductivity of the
nuclear pasta maintains a higher crust temperature during quiescence
and a layer of normal neutrons forms at the base of the
crust~\cite{Deibel:2016vbc}. Without neutrino heating, the cooling
observations of MXB~1659-29 are fit with a crust impurity of $Q_{\rm
imp} =1$ and a pasta impurity parameter of $Q_{\rm imp} = 20$. We
find that, similar to the model without nuclear pasta, as neutrino
heating is increased in the inner crust, the pasta impurity
parameter must decrease to reestablish a fit to the observations.

Note that the cooling of MXB~1659-29 may be fit with other neutron
star masses and radii~\cite{Brown:2009kw} and the results in
Fig.~\ref{Fig9} are for a fixed gravity in the neutron star crust
(and crust thickness). Therefore our studies above are qualitative
only as the amount of neutron heating for a $M = 1.6 \,
\mathrm{M_{\odot}}$ and $R = 11.2 \, \mathrm{km}$ neutron star from
$\alpha$-$Fe$ collision in particular is just about $0.13$ MeV and
is bigger only for more massive and/or compact stars (See also
Fig.\,\ref{Fig5}). Cooling light curve shapes are degenerate in
several parameters, for example: the neutron star gravity, the crust
impurity parameter, and the mass accretion rate. Because the effect
of neutrino heating is difficult to delineate from the effects of
other model parameters we therefore can not determine if neutrino
heating is present during outburst. It is worth noting, however,
that if deep crustal heating from neutrinos is present then existing
constraints derived from cooling light curves will need to be
revisited, likely requiring a higher crust thermal conductivity or a
different neutron star gravity.



\section{Conclusion and Outlook}
\label{Conclusions} We presented a new mechanism of deep crustal
heating of neutron stars in mass-transferring binaries by neutrinos
that are decay remnants of charged pions produced at the surface of
neutron stars. Our calculations showed that massive and compact
stars can accelerate infalling matter to energies substantially
larger than the pion-production threshold resulting in ample
generation of neutrinos. Approximately half of these neutrinos
travel into the inner crust and deposit
$\approx 1 \textrm{--} 2\, \mathrm{MeV}$ per accreted nucleon
for massive and compact stars.

The deep crustal heating from neutrinos is comparable in strength to
pycnonuclear fusion reactions and other non-equilibrium nuclear
reactions taking place during active accretion. Additional deep
crustal heating will affect the cooling light curves of accreting
neutron stars at late times $\gtrsim 300 \, \mathrm{days}$ into
quiescence. The effect is most pronounced when the star is massive
and might help distinguish high-mass stars from low-mass stars. In
general, for a fixed gravity in the neutron star crust we find that
additional deep crustal heating requires a higher thermal
conductivity for the crust and the crust impurity parameter must be
lowered. In the particular case of MXB~1659-29, for a model without
nuclear pasta and the S03 pairing gap, $Q_{\rm imp} \lesssim 3$ is
required if any neutrino heating is added. For a model with nuclear
pasta and the G08 pairing gap, $Q_{\rm imp} \lesssim 20$ for nuclear
pasta is needed if neutrino heating is present.

Our calculation of pion production assumes that the incoming protons
are slowed by Coulomb collisions with atmospheric
electrons~\cite{Zeldovich:1969}. Plasma instabilities or a
collisionless shock may instead stop the proton beam (\textit{e.g.}
see Ref. \cite{Shapiro:1975}), reducing the rate of nuclear
collisions. In addition, depending on the accretion geometry, the
incoming particles may not have the full free-fall velocity,
\textit{e.g.} in disk accretion if the disk reaches all the way to
the neutron star surface. Neutrino heating may operate only with a
quasi-spherical accretion flow or if the neutron star lies within
the last stable orbit (\textit{e.g.} see discussion in
Ref.~\cite{Bildsten:2003ApJ}).

There is also a strong sensitivity of our results to the pion
production cross sections at near threshold energies. Pion
production may play a significant role in stellar environments and
in particular, a better knowledge of pion production cross sections
in $p$-$p$, $p$-$Fe$, $\alpha$-$\alpha$, and $\alpha$-$Fe$ reactions
at beam energies $150$-$600$ MeV/nucleon may help to better
understand the structure and transport properties of neutron star
crusts from cooling observations.


\begin{acknowledgments}
\vspace{-0.4cm} We thank Professors Hans-Otto Meyer, Hendrik Schatz,
and Rex Tayloe for many helpful discussions. F.J.F., C.J.H., and
Z.L. are supported by the U.S. Department of Energy (DOE) grants
DE-FG02-87ER40365 (Indiana University), DE-SC0008808 (NUCLEI SciDAC
Collaboration) and by the National Science Foundation through XSEDE
resources provided by the National Institute for Computational
Sciences under grant TG-AST100014. E.F.B. is  supported by  the  US
National  Science Foundation under grant AST-1516969.  A.C. is
supported by an NSERC Discovery Grant and is a member of the Centre
de Recherche en Astrophysique du Quebec (CRAQ). B.A.L. is supported
by the U.S. Department of Energy, Office of Science, under Award
Number DE-SC0013702 (Texas A\&M University-Commerce) and the
National Natural Science Foundation of China under Grant No.
11320101004.
\end{acknowledgments}
\vfill

\bibliography{ReferencesFJF}

\begin{thebibliography}{69}%
\makeatletter
\providecommand \@ifxundefined [1]{%
 \@ifx{#1\undefined}
}%
\providecommand \@ifnum [1]{%
 \ifnum #1\expandafter \@firstoftwo
 \else \expandafter \@secondoftwo
 \fi
}%
\providecommand \@ifx [1]{%
 \ifx #1\expandafter \@firstoftwo
 \else \expandafter \@secondoftwo
 \fi
}%
\providecommand \natexlab [1]{#1}%
\providecommand \enquote  [1]{``#1''}%
\providecommand \bibnamefont  [1]{#1}%
\providecommand \bibfnamefont [1]{#1}%
\providecommand \citenamefont [1]{#1}%
\providecommand \href@noop [0]{\@secondoftwo}%
\providecommand \href [0]{\begingroup \@sanitize@url \@href}%
\providecommand \@href[1]{\@@startlink{#1}\@@href}%
\providecommand \@@href[1]{\endgroup#1\@@endlink}%
\providecommand \@sanitize@url [0]{\catcode `\\12\catcode `\$12\catcode
  `\&12\catcode `\#12\catcode `\^12\catcode `\_12\catcode `\%12\relax}%
\providecommand \@@startlink[1]{}%
\providecommand \@@endlink[0]{}%
\providecommand \url  [0]{\begingroup\@sanitize@url \@url }%
\providecommand \@url [1]{\endgroup\@href {#1}{\urlprefix }}%
\providecommand \urlprefix  [0]{URL }%
\providecommand \Eprint [0]{\href }%
\providecommand \doibase [0]{http://dx.doi.org/}%
\providecommand \selectlanguage [0]{\@gobble}%
\providecommand \bibinfo  [0]{\@secondoftwo}%
\providecommand \bibfield  [0]{\@secondoftwo}%
\providecommand \translation [1]{[#1]}%
\providecommand \BibitemOpen [0]{}%
\providecommand \bibitemStop [0]{}%
\providecommand \bibitemNoStop [0]{.\EOS\space}%
\providecommand \EOS [0]{\spacefactor3000\relax}%
\providecommand \BibitemShut  [1]{\csname bibitem#1\endcsname}%
\let\auto@bib@innerbib\@empty
\bibitem [{\citenamefont {{Bisnovaty{\u i}-Kogan}}\ and\ \citenamefont
  {{Chechetkin}}(1979)}]{bisnovatyi1979}%
  \BibitemOpen
  \bibfield  {author} {\bibinfo {author} {\bibfnamefont {G.~S.}\ \bibnamefont
  {{Bisnovaty{\u i}-Kogan}}}\ and\ \bibinfo {author} {\bibfnamefont {V.~M.}\
  \bibnamefont {{Chechetkin}}},\ }\href {\doibase
  10.1070/PU1979v022n02ABEH005418} {\bibfield  {journal} {\bibinfo  {journal}
  {Soviet Physics Uspekhi}\ }\textbf {\bibinfo {volume} {22}},\ \bibinfo
  {pages} {89} (\bibinfo {year} {1979})}\BibitemShut {NoStop}%
\bibitem [{\citenamefont {{Sato}}(1979)}]{sato1979}%
  \BibitemOpen
  \bibfield  {author} {\bibinfo {author} {\bibfnamefont {K.}~\bibnamefont
  {{Sato}}},\ }\href {\doibase 10.1143/PTP.62.957} {\bibfield  {journal}
  {\bibinfo  {journal} {Progress of Theoretical Physics}\ }\textbf {\bibinfo
  {volume} {62}},\ \bibinfo {pages} {957} (\bibinfo {year} {1979})}\BibitemShut
  {NoStop}%
\bibitem [{\citenamefont {{Haensel}}\ and\ \citenamefont
  {{Zdunik}}(1990)}]{haensel1990}%
  \BibitemOpen
  \bibfield  {author} {\bibinfo {author} {\bibfnamefont {P.}~\bibnamefont
  {{Haensel}}}\ and\ \bibinfo {author} {\bibfnamefont {J.~L.}\ \bibnamefont
  {{Zdunik}}},\ }\href@noop {} {\bibfield  {journal} {\bibinfo  {journal}
  {Astron. \& Astrophys.}\ }\textbf {\bibinfo {volume} {227}},\ \bibinfo
  {pages} {431} (\bibinfo {year} {1990})}\BibitemShut {NoStop}%
\bibitem [{\citenamefont {{Haensel}}\ and\ \citenamefont
  {{Zdunik}}(2003)}]{haensel2003}%
  \BibitemOpen
  \bibfield  {author} {\bibinfo {author} {\bibfnamefont {P.}~\bibnamefont
  {{Haensel}}}\ and\ \bibinfo {author} {\bibfnamefont {J.~L.}\ \bibnamefont
  {{Zdunik}}},\ }\href {\doibase 10.1051/0004-6361:20030708} {\bibfield
  {journal} {\bibinfo  {journal} {Astron. \& Astrophys.}\ }\textbf {\bibinfo
  {volume} {404}},\ \bibinfo {pages} {L33} (\bibinfo {year}
  {2003})}\BibitemShut {NoStop}%
\bibitem [{\citenamefont {Haensel}\ and\ \citenamefont
  {Zdunik}(2008)}]{Haensel:2007wy}%
  \BibitemOpen
  \bibfield  {author} {\bibinfo {author} {\bibfnamefont {P.}~\bibnamefont
  {Haensel}}\ and\ \bibinfo {author} {\bibfnamefont {J.~L.}\ \bibnamefont
  {Zdunik}},\ }\href {\doibase 10.1051/0004-6361:20078578} {\bibfield
  {journal} {\bibinfo  {journal} {Astron. Astrophys.}\ }\textbf {\bibinfo
  {volume} {480}},\ \bibinfo {pages} {459} (\bibinfo {year}
  {2008})}\BibitemShut {NoStop}%
\bibitem [{\citenamefont {Ushomirsky}\ and\ \citenamefont
  {Rutledge}(2001)}]{Ushomirsky:2001pd}%
  \BibitemOpen
  \bibfield  {author} {\bibinfo {author} {\bibfnamefont {G.}~\bibnamefont
  {Ushomirsky}}\ and\ \bibinfo {author} {\bibfnamefont {R.~E.}\ \bibnamefont
  {Rutledge}},\ }\href {\doibase 10.1046/j.1365-8711.2001.04515.x} {\bibfield
  {journal} {\bibinfo  {journal} {Mon. Not. Roy. Astron. Soc.}\ }\textbf
  {\bibinfo {volume} {325}},\ \bibinfo {pages} {1157} (\bibinfo {year}
  {2001})}\BibitemShut {NoStop}%
\bibitem [{\citenamefont {{Rutledge}}\ \emph {et~al.}(2002)\citenamefont
  {{Rutledge}}, \citenamefont {{Bildsten}}, \citenamefont {{Brown}},
  \citenamefont {{Pavlov}}, \citenamefont {{Zavlin}},\ and\ \citenamefont
  {{Ushomirsky}}}]{Rutledge:2002}%
  \BibitemOpen
  \bibfield  {author} {\bibinfo {author} {\bibfnamefont {R.~E.}\ \bibnamefont
  {{Rutledge}}}, \bibinfo {author} {\bibfnamefont {L.}~\bibnamefont
  {{Bildsten}}}, \bibinfo {author} {\bibfnamefont {E.~F.}\ \bibnamefont
  {{Brown}}}, \bibinfo {author} {\bibfnamefont {G.~G.}\ \bibnamefont
  {{Pavlov}}}, \bibinfo {author} {\bibfnamefont {V.~E.}\ \bibnamefont
  {{Zavlin}}}, \ and\ \bibinfo {author} {\bibfnamefont {G.}~\bibnamefont
  {{Ushomirsky}}},\ }\href {\doibase 10.1086/342745} {\bibfield  {journal}
  {\bibinfo  {journal} {\apj}\ }\textbf {\bibinfo {volume} {580}},\ \bibinfo
  {pages} {413} (\bibinfo {year} {2002})}\BibitemShut {NoStop}%
\bibitem [{\citenamefont {{Shternin}}\ \emph {et~al.}(2007)\citenamefont
  {{Shternin}}, \citenamefont {{Yakovlev}}, \citenamefont {{Haensel}},\ and\
  \citenamefont {{Potekhin}}}]{shternin2007}%
  \BibitemOpen
  \bibfield  {author} {\bibinfo {author} {\bibfnamefont {P.~S.}\ \bibnamefont
  {{Shternin}}}, \bibinfo {author} {\bibfnamefont {D.~G.}\ \bibnamefont
  {{Yakovlev}}}, \bibinfo {author} {\bibfnamefont {P.}~\bibnamefont
  {{Haensel}}}, \ and\ \bibinfo {author} {\bibfnamefont {A.~Y.}\ \bibnamefont
  {{Potekhin}}},\ }\href {\doibase 10.1111/j.1745-3933.2007.00386.x} {\bibfield
   {journal} {\bibinfo  {journal} {Mon. Not. Roy. Astron. Soc.}\ }\textbf
  {\bibinfo {volume} {382}},\ \bibinfo {pages} {L43} (\bibinfo {year}
  {2007})}\BibitemShut {NoStop}%
\bibitem [{\citenamefont {Brown}\ and\ \citenamefont
  {Cumming}(2009)}]{Brown:2009kw}%
  \BibitemOpen
  \bibfield  {author} {\bibinfo {author} {\bibfnamefont {E.~F.}\ \bibnamefont
  {Brown}}\ and\ \bibinfo {author} {\bibfnamefont {A.}~\bibnamefont
  {Cumming}},\ }\href {\doibase 10.1088/0004-637X/698/2/1020} {\bibfield
  {journal} {\bibinfo  {journal} {Astrophys. J.}\ }\textbf {\bibinfo {volume}
  {698}},\ \bibinfo {pages} {1020} (\bibinfo {year} {2009})}\BibitemShut
  {NoStop}%
\bibitem [{\citenamefont {Cumming}\ \emph {et~al.}(2006)\citenamefont
  {Cumming}, \citenamefont {Macbeth}, \citenamefont {in~'t Zand},\ and\
  \citenamefont {Page}}]{Cumming:2005kk}%
  \BibitemOpen
  \bibfield  {author} {\bibinfo {author} {\bibfnamefont {A.}~\bibnamefont
  {Cumming}}, \bibinfo {author} {\bibfnamefont {J.}~\bibnamefont {Macbeth}},
  \bibinfo {author} {\bibfnamefont {J.~J.~M.}\ \bibnamefont {in~'t Zand}}, \
  and\ \bibinfo {author} {\bibfnamefont {D.}~\bibnamefont {Page}},\ }\href
  {\doibase 10.1086/504698} {\bibfield  {journal} {\bibinfo  {journal}
  {Astrophys. J.}\ }\textbf {\bibinfo {volume} {646}},\ \bibinfo {pages} {429}
  (\bibinfo {year} {2006})}\BibitemShut {NoStop}%
\bibitem [{\citenamefont {Deibel}\ \emph {et~al.}(2015)\citenamefont {Deibel},
  \citenamefont {Cumming}, \citenamefont {Brown},\ and\ \citenamefont
  {Page}}]{Deibel:2015ola}%
  \BibitemOpen
  \bibfield  {author} {\bibinfo {author} {\bibfnamefont {A.}~\bibnamefont
  {Deibel}}, \bibinfo {author} {\bibfnamefont {A.}~\bibnamefont {Cumming}},
  \bibinfo {author} {\bibfnamefont {E.~F.}\ \bibnamefont {Brown}}, \ and\
  \bibinfo {author} {\bibfnamefont {D.}~\bibnamefont {Page}},\ }\href {\doibase
  10.1088/2041-8205/809/2/L31} {\bibfield  {journal} {\bibinfo  {journal}
  {Astrophys. J.}\ }\textbf {\bibinfo {volume} {809}},\ \bibinfo {pages} {L31}
  (\bibinfo {year} {2015})}\BibitemShut {NoStop}%
\bibitem [{\citenamefont {Bildsten}\ \emph {et~al.}(1992)\citenamefont
  {Bildsten}, \citenamefont {Salpeter},\ and\ \citenamefont
  {Wasserman}}]{Bildsten:1992ApJ}%
  \BibitemOpen
  \bibfield  {author} {\bibinfo {author} {\bibfnamefont {L.}~\bibnamefont
  {Bildsten}}, \bibinfo {author} {\bibfnamefont {E.~E.}\ \bibnamefont
  {Salpeter}}, \ and\ \bibinfo {author} {\bibfnamefont {I.}~\bibnamefont
  {Wasserman}},\ }\href {\doibase 10.1086/170860} {\bibfield  {journal}
  {\bibinfo  {journal} {Astrophys. J.}\ }\textbf {\bibinfo {volume} {384}},\
  \bibinfo {pages} {143} (\bibinfo {year} {1992})}\BibitemShut {NoStop}%
\bibitem [{\citenamefont {Scholberg}(2006)}]{Scholberg:2005qs}%
  \BibitemOpen
  \bibfield  {author} {\bibinfo {author} {\bibfnamefont {K.}~\bibnamefont
  {Scholberg}},\ }\href {\doibase 10.1103/PhysRevD.73.033005} {\bibfield
  {journal} {\bibinfo  {journal} {Phys. Rev.}\ }\textbf {\bibinfo {volume}
  {D73}},\ \bibinfo {pages} {033005} (\bibinfo {year} {2006})}\BibitemShut
  {NoStop}%
\bibitem [{\citenamefont {Amsler}\ \emph {et~al.}(2008)\citenamefont {Amsler}
  \emph {et~al.}}]{Amsler:2008zzb}%
  \BibitemOpen
  \bibfield  {author} {\bibinfo {author} {\bibfnamefont {C.}~\bibnamefont
  {Amsler}} \emph {et~al.} (\bibinfo {collaboration} {Particle Data Group}),\
  }\href {\doibase 10.1016/j.physletb.2008.07.018} {\bibfield  {journal}
  {\bibinfo  {journal} {Phys. Lett.}\ }\textbf {\bibinfo {volume} {B667}},\
  \bibinfo {pages} {1} (\bibinfo {year} {2008})}\BibitemShut {NoStop}%
\bibitem [{\citenamefont {Tavernier}(2010)}]{Tavernier:2010}%
  \BibitemOpen
  \bibfield  {author} {\bibinfo {author} {\bibfnamefont {S.}~\bibnamefont
  {Tavernier}},\ }\enquote {\bibinfo {title} {Experimental techniques in
  nuclear and particle physics},}\ \ (\bibinfo  {publisher} {Springer-Verlag
  Berlin Heidelberg},\ \bibinfo {year} {2010})\BibitemShut {NoStop}%
\bibitem [{\citenamefont {Horowitz}(2002)}]{Horowitz:2001xf}%
  \BibitemOpen
  \bibfield  {author} {\bibinfo {author} {\bibfnamefont {C.~J.}\ \bibnamefont
  {Horowitz}},\ }\href {\doibase 10.1103/PhysRevD.65.043001} {\bibfield
  {journal} {\bibinfo  {journal} {Phys. Rev.}\ }\textbf {\bibinfo {volume}
  {D65}},\ \bibinfo {pages} {043001} (\bibinfo {year} {2002})}\BibitemShut
  {NoStop}%
\bibitem [{\citenamefont {Horowitz}\ \emph {et~al.}(2003)\citenamefont
  {Horowitz}, \citenamefont {Coakley},\ and\ \citenamefont
  {McKinsey}}]{Horowitz:2003cz}%
  \BibitemOpen
  \bibfield  {author} {\bibinfo {author} {\bibfnamefont {C.~J.}\ \bibnamefont
  {Horowitz}}, \bibinfo {author} {\bibfnamefont {K.~J.}\ \bibnamefont
  {Coakley}}, \ and\ \bibinfo {author} {\bibfnamefont {D.~N.}\ \bibnamefont
  {McKinsey}},\ }\href {\doibase 10.1103/PhysRevD.68.023005} {\bibfield
  {journal} {\bibinfo  {journal} {Phys. Rev.}\ }\textbf {\bibinfo {volume}
  {D68}},\ \bibinfo {pages} {023005} (\bibinfo {year} {2003})}\BibitemShut
  {NoStop}%
\bibitem [{\citenamefont {Bahcall}(1964)}]{Bahcall:1964zzb}%
  \BibitemOpen
  \bibfield  {author} {\bibinfo {author} {\bibfnamefont {J.~N.}\ \bibnamefont
  {Bahcall}},\ }\href {\doibase 10.1103/PhysRev.136.B1164} {\bibfield
  {journal} {\bibinfo  {journal} {Phys. Rev.}\ }\textbf {\bibinfo {volume}
  {136}},\ \bibinfo {pages} {B1164} (\bibinfo {year} {1964})}\BibitemShut
  {NoStop}%
\bibitem [{\citenamefont {Langanke}\ \emph {et~al.}(2004)\citenamefont
  {Langanke}, \citenamefont {Martinez-Pinedo}, \citenamefont {von
  Neumann-Cosel},\ and\ \citenamefont {Richter}}]{Langanke:2004vx}%
  \BibitemOpen
  \bibfield  {author} {\bibinfo {author} {\bibfnamefont {K.}~\bibnamefont
  {Langanke}}, \bibinfo {author} {\bibfnamefont {G.}~\bibnamefont
  {Martinez-Pinedo}}, \bibinfo {author} {\bibfnamefont {P.}~\bibnamefont {von
  Neumann-Cosel}}, \ and\ \bibinfo {author} {\bibfnamefont {A.}~\bibnamefont
  {Richter}},\ }\href {\doibase 10.1103/PhysRevLett.93.202501} {\bibfield
  {journal} {\bibinfo  {journal} {Phys. Rev. Lett.}\ }\textbf {\bibinfo
  {volume} {93}},\ \bibinfo {pages} {202501} (\bibinfo {year}
  {2004})}\BibitemShut {NoStop}%
\bibitem [{\citenamefont {Haensel}\ \emph {et~al.}(1989)\citenamefont
  {Haensel}, \citenamefont {Zdunik},\ and\ \citenamefont
  {Dobaczewski}}]{Haensel:1989}%
  \BibitemOpen
  \bibfield  {author} {\bibinfo {author} {\bibfnamefont {P.}~\bibnamefont
  {Haensel}}, \bibinfo {author} {\bibfnamefont {J.~L.}\ \bibnamefont {Zdunik}},
  \ and\ \bibinfo {author} {\bibfnamefont {J.}~\bibnamefont {Dobaczewski}},\
  }\href@noop {} {\bibfield  {journal} {\bibinfo  {journal} {Astron.
  Astrophys.}\ }\textbf {\bibinfo {volume} {222}},\ \bibinfo {pages} {353}
  (\bibinfo {year} {1989})}\BibitemShut {NoStop}%
\bibitem [{\citenamefont {Ravenhall}\ \emph {et~al.}(1983)\citenamefont
  {Ravenhall}, \citenamefont {Pethick},\ and\ \citenamefont
  {Wilson}}]{Ravenhall:1983uh}%
  \BibitemOpen
  \bibfield  {author} {\bibinfo {author} {\bibfnamefont {D.~G.}\ \bibnamefont
  {Ravenhall}}, \bibinfo {author} {\bibfnamefont {C.~J.}\ \bibnamefont
  {Pethick}}, \ and\ \bibinfo {author} {\bibfnamefont {J.~R.}\ \bibnamefont
  {Wilson}},\ }\href@noop {} {\bibfield  {journal} {\bibinfo  {journal} {Phys.
  Rev. Lett.}\ }\textbf {\bibinfo {volume} {50}},\ \bibinfo {pages} {2066}
  (\bibinfo {year} {1983})}\BibitemShut {NoStop}%
\bibitem [{\citenamefont {Hashimoto}\ \emph {et~al.}(1984)\citenamefont
  {Hashimoto}, \citenamefont {Seki},\ and\ \citenamefont
  {Yamada}}]{Hashimoto:1984}%
  \BibitemOpen
  \bibfield  {author} {\bibinfo {author} {\bibfnamefont {M.}~\bibnamefont
  {Hashimoto}}, \bibinfo {author} {\bibfnamefont {H.}~\bibnamefont {Seki}}, \
  and\ \bibinfo {author} {\bibfnamefont {M.}~\bibnamefont {Yamada}},\
  }\href@noop {} {\bibfield  {journal} {\bibinfo  {journal} {Prog. Theor.
  Phys.}\ }\textbf {\bibinfo {volume} {71}},\ \bibinfo {pages} {320} (\bibinfo
  {year} {1984})}\BibitemShut {NoStop}%
\bibitem [{\citenamefont {Lorenz}\ \emph {et~al.}(1993)\citenamefont {Lorenz},
  \citenamefont {Ravenhall},\ and\ \citenamefont {Pethick}}]{Lorenz:1992zz}%
  \BibitemOpen
  \bibfield  {author} {\bibinfo {author} {\bibfnamefont {C.~P.}\ \bibnamefont
  {Lorenz}}, \bibinfo {author} {\bibfnamefont {D.~G.}\ \bibnamefont
  {Ravenhall}}, \ and\ \bibinfo {author} {\bibfnamefont {C.~J.}\ \bibnamefont
  {Pethick}},\ }\href {\doibase 10.1103/PhysRevLett.70.379} {\bibfield
  {journal} {\bibinfo  {journal} {Phys. Rev. Lett.}\ }\textbf {\bibinfo
  {volume} {70}},\ \bibinfo {pages} {379} (\bibinfo {year} {1993})}\BibitemShut
  {NoStop}%
\bibitem [{\citenamefont {Negele}\ and\ \citenamefont
  {Vautherin}(1973)}]{Negele:1971vb}%
  \BibitemOpen
  \bibfield  {author} {\bibinfo {author} {\bibfnamefont {J.~W.}\ \bibnamefont
  {Negele}}\ and\ \bibinfo {author} {\bibfnamefont {D.}~\bibnamefont
  {Vautherin}},\ }\href@noop {} {\bibfield  {journal} {\bibinfo  {journal}
  {Nucl. Phys.}\ }\textbf {\bibinfo {volume} {A207}},\ \bibinfo {pages} {298}
  (\bibinfo {year} {1973})}\BibitemShut {NoStop}%
\bibitem [{\citenamefont {Chen}\ and\ \citenamefont
  {Piekarewicz}(2014)}]{Chen:2014sca}%
  \BibitemOpen
  \bibfield  {author} {\bibinfo {author} {\bibfnamefont {W.-C.}\ \bibnamefont
  {Chen}}\ and\ \bibinfo {author} {\bibfnamefont {J.}~\bibnamefont
  {Piekarewicz}},\ }\href {\doibase 10.1103/PhysRevC.90.044305} {\bibfield
  {journal} {\bibinfo  {journal} {Phys. Rev.}\ }\textbf {\bibinfo {volume}
  {C90}},\ \bibinfo {pages} {044305} (\bibinfo {year} {2014})}\BibitemShut
  {NoStop}%
\bibitem [{\citenamefont {Demorest}\ \emph {et~al.}(2010)\citenamefont
  {Demorest}, \citenamefont {Pennucci}, \citenamefont {Ransom}, \citenamefont
  {Roberts},\ and\ \citenamefont {Hessels}}]{Demorest:2010bx}%
  \BibitemOpen
  \bibfield  {author} {\bibinfo {author} {\bibfnamefont {P.~B.}\ \bibnamefont
  {Demorest}}, \bibinfo {author} {\bibfnamefont {T.}~\bibnamefont {Pennucci}},
  \bibinfo {author} {\bibfnamefont {S.~M.}\ \bibnamefont {Ransom}}, \bibinfo
  {author} {\bibfnamefont {M.~S.~E.}\ \bibnamefont {Roberts}}, \ and\ \bibinfo
  {author} {\bibfnamefont {J.~W.~T.}\ \bibnamefont {Hessels}},\ }\href
  {\doibase 10.1038/nature09466} {\bibfield  {journal} {\bibinfo  {journal}
  {Nature}\ }\textbf {\bibinfo {volume} {467}},\ \bibinfo {pages} {1081}
  (\bibinfo {year} {2010})}\BibitemShut {NoStop}%
\bibitem [{\citenamefont {Antoniadis}\ \emph {et~al.}(2013)\citenamefont
  {Antoniadis} \emph {et~al.}}]{Antoniadis:2013pzd}%
  \BibitemOpen
  \bibfield  {author} {\bibinfo {author} {\bibfnamefont {J.}~\bibnamefont
  {Antoniadis}} \emph {et~al.},\ }\href {\doibase 10.1126/science.1233232}
  {\bibfield  {journal} {\bibinfo  {journal} {Science}\ }\textbf {\bibinfo
  {volume} {340}},\ \bibinfo {pages} {6131} (\bibinfo {year}
  {2013})}\BibitemShut {NoStop}%
\bibitem [{\citenamefont {Fonseca}\ \emph {et~al.}(2016)\citenamefont {Fonseca}
  \emph {et~al.}}]{Fonseca:2016tux}%
  \BibitemOpen
  \bibfield  {author} {\bibinfo {author} {\bibfnamefont {E.}~\bibnamefont
  {Fonseca}} \emph {et~al.},\ }\href {\doibase 10.3847/0004-637X/832/2/167}
  {\bibfield  {journal} {\bibinfo  {journal} {Astrophys. J.}\ }\textbf
  {\bibinfo {volume} {832}},\ \bibinfo {pages} {167} (\bibinfo {year}
  {2016})}\BibitemShut {NoStop}%
\bibitem [{\citenamefont {Horowitz}\ and\ \citenamefont
  {Piekarewicz}(2001)}]{Horowitz:2000xj}%
  \BibitemOpen
  \bibfield  {author} {\bibinfo {author} {\bibfnamefont {C.~J.}\ \bibnamefont
  {Horowitz}}\ and\ \bibinfo {author} {\bibfnamefont {J.}~\bibnamefont
  {Piekarewicz}},\ }\href@noop {} {\bibfield  {journal} {\bibinfo  {journal}
  {Phys. Rev. Lett.}\ }\textbf {\bibinfo {volume} {86}},\ \bibinfo {pages}
  {5647} (\bibinfo {year} {2001})}\BibitemShut {NoStop}%
\bibitem [{\citenamefont {Fattoyev}\ \emph {et~al.}(2012)\citenamefont
  {Fattoyev}, \citenamefont {Newton}, \citenamefont {Xu},\ and\ \citenamefont
  {Li}}]{Fattoyev:2012ch}%
  \BibitemOpen
  \bibfield  {author} {\bibinfo {author} {\bibfnamefont {F.}~\bibnamefont
  {Fattoyev}}, \bibinfo {author} {\bibfnamefont {W.}~\bibnamefont {Newton}},
  \bibinfo {author} {\bibfnamefont {J.}~\bibnamefont {Xu}}, \ and\ \bibinfo
  {author} {\bibfnamefont {B.-A.}\ \bibnamefont {Li}},\ }\href {\doibase
  10.1103/PhysRevC.86.025804} {\bibfield  {journal} {\bibinfo  {journal} {Phys.
  Rev.}\ }\textbf {\bibinfo {volume} {C86}},\ \bibinfo {pages} {025804}
  (\bibinfo {year} {2012})}\BibitemShut {NoStop}%
\bibitem [{\citenamefont {Hebeler}\ \emph {et~al.}(2010)\citenamefont
  {Hebeler}, \citenamefont {Lattimer}, \citenamefont {Pethick},\ and\
  \citenamefont {Schwenk}}]{Hebeler:2010jx}%
  \BibitemOpen
  \bibfield  {author} {\bibinfo {author} {\bibfnamefont {K.}~\bibnamefont
  {Hebeler}}, \bibinfo {author} {\bibfnamefont {J.}~\bibnamefont {Lattimer}},
  \bibinfo {author} {\bibfnamefont {C.}~\bibnamefont {Pethick}}, \ and\
  \bibinfo {author} {\bibfnamefont {A.}~\bibnamefont {Schwenk}},\ }\href
  {\doibase 10.1103/PhysRevLett.105.161102} {\bibfield  {journal} {\bibinfo
  {journal} {Phys. Rev. Lett.}\ }\textbf {\bibinfo {volume} {105}},\ \bibinfo
  {pages} {161102} (\bibinfo {year} {2010})}\BibitemShut {NoStop}%
\bibitem [{\citenamefont {Casares}\ \emph {et~al.}()\citenamefont {Casares},
  \citenamefont {Jonker},\ and\ \citenamefont {Israelian}}]{Casares:2017jah}%
  \BibitemOpen
  \bibfield  {author} {\bibinfo {author} {\bibfnamefont {J.}~\bibnamefont
  {Casares}}, \bibinfo {author} {\bibfnamefont {P.~G.}\ \bibnamefont {Jonker}},
  \ and\ \bibinfo {author} {\bibfnamefont {G.}~\bibnamefont {Israelian}},\
  }\href@noop {} {\ }\Eprint {http://arxiv.org/abs/1701.07450}
  {arXiv:1701.07450} \BibitemShut {NoStop}%
\bibitem [{\citenamefont {Chang}\ \emph {et~al.}(2010)\citenamefont {Chang},
  \citenamefont {Bildsten},\ and\ \citenamefont {Arras}}]{Chang:2010sx}%
  \BibitemOpen
  \bibfield  {author} {\bibinfo {author} {\bibfnamefont {P.}~\bibnamefont
  {Chang}}, \bibinfo {author} {\bibfnamefont {L.}~\bibnamefont {Bildsten}}, \
  and\ \bibinfo {author} {\bibfnamefont {P.}~\bibnamefont {Arras}},\ }\href
  {\doibase 10.1088/0004-637X/723/1/719} {\bibfield  {journal} {\bibinfo
  {journal} {Astrophys. J.}\ }\textbf {\bibinfo {volume} {723}},\ \bibinfo
  {pages} {719} (\bibinfo {year} {2010})}\BibitemShut {NoStop}%
\bibitem [{\citenamefont {in~'t Zand}\ \emph {et~al.}(2005)\citenamefont {in~'t
  Zand}, \citenamefont {Cumming}, \citenamefont {van~der Sluys}, \citenamefont
  {Verbunt},\ and\ \citenamefont {Pols}}]{intZand:2005sfk}%
  \BibitemOpen
  \bibfield  {author} {\bibinfo {author} {\bibfnamefont {J.~J.~M.}\
  \bibnamefont {in~'t Zand}}, \bibinfo {author} {\bibfnamefont
  {A.}~\bibnamefont {Cumming}}, \bibinfo {author} {\bibfnamefont {M.~V.}\
  \bibnamefont {van~der Sluys}}, \bibinfo {author} {\bibfnamefont
  {F.}~\bibnamefont {Verbunt}}, \ and\ \bibinfo {author} {\bibfnamefont
  {O.~R.}\ \bibnamefont {Pols}},\ }\href {\doibase 10.1051/0004-6361:20053002}
  {\bibfield  {journal} {\bibinfo  {journal} {Astron. Astrophys.}\ }\textbf
  {\bibinfo {volume} {441}},\ \bibinfo {pages} {675 } (\bibinfo {year}
  {2005})}\BibitemShut {NoStop}%
\bibitem [{\citenamefont {Crawford}\ \emph {et~al.}(1980)\citenamefont
  {Crawford} \emph {et~al.}}]{Crawford:1979ia}%
  \BibitemOpen
  \bibfield  {author} {\bibinfo {author} {\bibfnamefont {J.~F.}\ \bibnamefont
  {Crawford}} \emph {et~al.},\ }\href {\doibase 10.1103/PhysRevC.22.1184}
  {\bibfield  {journal} {\bibinfo  {journal} {Phys. Rev.}\ }\textbf {\bibinfo
  {volume} {C22}},\ \bibinfo {pages} {1184} (\bibinfo {year}
  {1980})}\BibitemShut {NoStop}%
\bibitem [{\citenamefont {Cochran}\ \emph {et~al.}(1972)\citenamefont
  {Cochran}, \citenamefont {Dean}, \citenamefont {Gram}, \citenamefont {Knapp},
  \citenamefont {Martin}, \citenamefont {Nagle}, \citenamefont {Perkins},
  \citenamefont {Shlaer}, \citenamefont {Thiessen},\ and\ \citenamefont
  {Theriot}}]{Cochran:1972pq}%
  \BibitemOpen
  \bibfield  {author} {\bibinfo {author} {\bibfnamefont {D.~R.~F.}\
  \bibnamefont {Cochran}}, \bibinfo {author} {\bibfnamefont {P.~N.}\
  \bibnamefont {Dean}}, \bibinfo {author} {\bibfnamefont {P.~A.~M.}\
  \bibnamefont {Gram}}, \bibinfo {author} {\bibfnamefont {E.~A.}\ \bibnamefont
  {Knapp}}, \bibinfo {author} {\bibfnamefont {E.~R.}\ \bibnamefont {Martin}},
  \bibinfo {author} {\bibfnamefont {D.~E.}\ \bibnamefont {Nagle}}, \bibinfo
  {author} {\bibfnamefont {R.~B.}\ \bibnamefont {Perkins}}, \bibinfo {author}
  {\bibfnamefont {W.~J.}\ \bibnamefont {Shlaer}}, \bibinfo {author}
  {\bibfnamefont {H.~A.}\ \bibnamefont {Thiessen}}, \ and\ \bibinfo {author}
  {\bibfnamefont {E.~D.}\ \bibnamefont {Theriot}},\ }\href {\doibase
  10.1103/PhysRevD.6.3085, 10.1103/PhysRevD.9.835} {\bibfield  {journal}
  {\bibinfo  {journal} {Phys. Rev.}\ }\textbf {\bibinfo {volume} {D6}},\
  \bibinfo {pages} {3085} (\bibinfo {year} {1972})},\ \bibinfo {note}
  {[Erratum: Phys. Rev.D9,835(1974)]}\BibitemShut {NoStop}%
\bibitem [{\citenamefont {Denes}\ \emph {et~al.}(1983)\citenamefont {Denes},
  \citenamefont {Dieterle}, \citenamefont {Wolfe}, \citenamefont {Bowles},
  \citenamefont {Dombeck}, \citenamefont {Simmons}, \citenamefont {Bhatia},
  \citenamefont {Glass},\ and\ \citenamefont {Tippens}}]{Denes:1983mw}%
  \BibitemOpen
  \bibfield  {author} {\bibinfo {author} {\bibfnamefont {P.}~\bibnamefont
  {Denes}}, \bibinfo {author} {\bibfnamefont {B.~D.}\ \bibnamefont {Dieterle}},
  \bibinfo {author} {\bibfnamefont {D.~M.}\ \bibnamefont {Wolfe}}, \bibinfo
  {author} {\bibfnamefont {T.}~\bibnamefont {Bowles}}, \bibinfo {author}
  {\bibfnamefont {T.}~\bibnamefont {Dombeck}}, \bibinfo {author} {\bibfnamefont
  {J.~E.}\ \bibnamefont {Simmons}}, \bibinfo {author} {\bibfnamefont {T.~S.}\
  \bibnamefont {Bhatia}}, \bibinfo {author} {\bibfnamefont {G.}~\bibnamefont
  {Glass}}, \ and\ \bibinfo {author} {\bibfnamefont {W.~B.}\ \bibnamefont
  {Tippens}},\ }\href {\doibase 10.1103/PhysRevC.27.1339} {\bibfield  {journal}
  {\bibinfo  {journal} {Phys. Rev.}\ }\textbf {\bibinfo {volume} {C27}},\
  \bibinfo {pages} {1339} (\bibinfo {year} {1983})}\BibitemShut {NoStop}%
\bibitem [{\citenamefont {Lemaire}\ \emph {et~al.}(1991)\citenamefont {Lemaire}
  \emph {et~al.}}]{Lemaire:1991xs}%
  \BibitemOpen
  \bibfield  {author} {\bibinfo {author} {\bibfnamefont {M.~C.}\ \bibnamefont
  {Lemaire}} \emph {et~al.},\ }in\ \href@noop {} {\emph {\bibinfo {booktitle}
  {{19th International Workshop on Gross Properties of Nuclei and Nuclear
  Excitations Hirschegg, Austria, January 21-26, 1991}}}}\ (\bibinfo {year}
  {1991})\ pp.\ \bibinfo {pages} {73--81}\BibitemShut {NoStop}%
\bibitem [{\citenamefont {Burman}\ \emph {et~al.}(1990)\citenamefont {Burman},
  \citenamefont {Potter},\ and\ \citenamefont {Smith}}]{Burman:1989dq}%
  \BibitemOpen
  \bibfield  {author} {\bibinfo {author} {\bibfnamefont {R.~L.}\ \bibnamefont
  {Burman}}, \bibinfo {author} {\bibfnamefont {M.~E.}\ \bibnamefont {Potter}},
  \ and\ \bibinfo {author} {\bibfnamefont {E.~S.}\ \bibnamefont {Smith}},\
  }\href {\doibase 10.1016/0168-9002(90)90012-U} {\bibfield  {journal}
  {\bibinfo  {journal} {Nucl. Instrum. Meth.}\ }\textbf {\bibinfo {volume}
  {A291}},\ \bibinfo {pages} {621} (\bibinfo {year} {1990})}\BibitemShut
  {NoStop}%
\bibitem [{\citenamefont {Daehnick}\ \emph {et~al.}(1995)\citenamefont
  {Daehnick} \emph {et~al.}}]{Daehnick:1995gw}%
  \BibitemOpen
  \bibfield  {author} {\bibinfo {author} {\bibfnamefont {W.~W.}\ \bibnamefont
  {Daehnick}} \emph {et~al.},\ }\href {\doibase 10.1103/PhysRevLett.74.2913}
  {\bibfield  {journal} {\bibinfo  {journal} {Phys. Rev. Lett.}\ }\textbf
  {\bibinfo {volume} {74}},\ \bibinfo {pages} {2913} (\bibinfo {year}
  {1995})}\BibitemShut {NoStop}%
\bibitem [{\citenamefont {Hardie}\ \emph {et~al.}(1997)\citenamefont {Hardie}
  \emph {et~al.}}]{Hardie:1997mg}%
  \BibitemOpen
  \bibfield  {author} {\bibinfo {author} {\bibfnamefont {J.~G.}\ \bibnamefont
  {Hardie}} \emph {et~al.},\ }\href {\doibase 10.1103/PhysRevC.56.20}
  {\bibfield  {journal} {\bibinfo  {journal} {Phys. Rev.}\ }\textbf {\bibinfo
  {volume} {C56}},\ \bibinfo {pages} {20} (\bibinfo {year} {1997})}\BibitemShut
  {NoStop}%
\bibitem [{\citenamefont {Flammang}\ \emph {et~al.}(1998)\citenamefont
  {Flammang} \emph {et~al.}}]{Flammang:1998bb}%
  \BibitemOpen
  \bibfield  {author} {\bibinfo {author} {\bibfnamefont {R.~W.}\ \bibnamefont
  {Flammang}} \emph {et~al.},\ }\href {\doibase 10.1103/PhysRevC.58.916}
  {\bibfield  {journal} {\bibinfo  {journal} {Phys. Rev.}\ }\textbf {\bibinfo
  {volume} {C58}},\ \bibinfo {pages} {916} (\bibinfo {year}
  {1998})}\BibitemShut {NoStop}%
\bibitem [{\citenamefont {Schwaller}\ \emph {et~al.}(1979)\citenamefont
  {Schwaller}, \citenamefont {Pepin}, \citenamefont {Favier}, \citenamefont
  {Richard-Serre}, \citenamefont {Measday},\ and\ \citenamefont
  {Renberg}}]{Schwaller:1979eu}%
  \BibitemOpen
  \bibfield  {author} {\bibinfo {author} {\bibfnamefont {P.}~\bibnamefont
  {Schwaller}}, \bibinfo {author} {\bibfnamefont {M.}~\bibnamefont {Pepin}},
  \bibinfo {author} {\bibfnamefont {B.}~\bibnamefont {Favier}}, \bibinfo
  {author} {\bibfnamefont {C.}~\bibnamefont {Richard-Serre}}, \bibinfo {author}
  {\bibfnamefont {D.~F.}\ \bibnamefont {Measday}}, \ and\ \bibinfo {author}
  {\bibfnamefont {P.~U.}\ \bibnamefont {Renberg}},\ }\href {\doibase
  10.1016/0375-9474(79)90040-X} {\bibfield  {journal} {\bibinfo  {journal}
  {Nucl. Phys.}\ }\textbf {\bibinfo {volume} {A316}},\ \bibinfo {pages} {317}
  (\bibinfo {year} {1979})}\BibitemShut {NoStop}%
\bibitem [{\citenamefont {Tsang}\ \emph {et~al.}(2017)\citenamefont {Tsang}
  \emph {et~al.}}]{Tsang:2016foy}%
  \BibitemOpen
  \bibfield  {author} {\bibinfo {author} {\bibfnamefont {M.~B.}\ \bibnamefont
  {Tsang}} \emph {et~al.} (\bibinfo {collaboration} {SpRIT}),\ }\href {\doibase
  10.1103/PhysRevC.95.044614} {\bibfield  {journal} {\bibinfo  {journal} {Phys.
  Rev.}\ }\textbf {\bibinfo {volume} {C95}},\ \bibinfo {pages} {044614}
  (\bibinfo {year} {2017})}\BibitemShut {NoStop}%
\bibitem [{\citenamefont {Bertsch}(1977)}]{Bertsch:1977ps}%
  \BibitemOpen
  \bibfield  {author} {\bibinfo {author} {\bibfnamefont {G.~F.}\ \bibnamefont
  {Bertsch}},\ }\href {\doibase 10.1103/PhysRevC.15.713} {\bibfield  {journal}
  {\bibinfo  {journal} {Phys. Rev.}\ }\textbf {\bibinfo {volume} {C15}},\
  \bibinfo {pages} {713} (\bibinfo {year} {1977})}\BibitemShut {NoStop}%
\bibitem [{\citenamefont {Sandel}\ \emph {et~al.}(1979)\citenamefont {Sandel},
  \citenamefont {Vary},\ and\ \citenamefont {Garpman}}]{Sandel:1979tg}%
  \BibitemOpen
  \bibfield  {author} {\bibinfo {author} {\bibfnamefont {M.}~\bibnamefont
  {Sandel}}, \bibinfo {author} {\bibfnamefont {J.~P.}\ \bibnamefont {Vary}}, \
  and\ \bibinfo {author} {\bibfnamefont {S.~I.~A.}\ \bibnamefont {Garpman}},\
  }\href {\doibase 10.1103/PhysRevC.20.744} {\bibfield  {journal} {\bibinfo
  {journal} {Phys. Rev.}\ }\textbf {\bibinfo {volume} {C20}},\ \bibinfo {pages}
  {744} (\bibinfo {year} {1979})}\BibitemShut {NoStop}%
\bibitem [{\citenamefont {Miller}\ \emph {et~al.}(1993)\citenamefont {Miller}
  \emph {et~al.}}]{Miller:1993zw}%
  \BibitemOpen
  \bibfield  {author} {\bibinfo {author} {\bibfnamefont {J.}~\bibnamefont
  {Miller}} \emph {et~al.},\ }\href {\doibase 10.1016/0370-2693(93)91314-D}
  {\bibfield  {journal} {\bibinfo  {journal} {Phys. Lett.}\ }\textbf {\bibinfo
  {volume} {B314}},\ \bibinfo {pages} {7} (\bibinfo {year} {1993})}\BibitemShut
  {NoStop}%
\bibitem [{\citenamefont {Bildsten}\ \emph {et~al.}(1993)\citenamefont
  {Bildsten}, \citenamefont {Salpeter},\ and\ \citenamefont
  {Wasserman}}]{Bildsten:1993ApJ}%
  \BibitemOpen
  \bibfield  {author} {\bibinfo {author} {\bibfnamefont {L.}~\bibnamefont
  {Bildsten}}, \bibinfo {author} {\bibfnamefont {E.~E.}\ \bibnamefont
  {Salpeter}}, \ and\ \bibinfo {author} {\bibfnamefont {I.}~\bibnamefont
  {Wasserman}},\ }\href {\doibase 10.1086/172621} {\bibfield  {journal}
  {\bibinfo  {journal} {Astrophys. J.}\ }\textbf {\bibinfo {volume} {408}},\
  \bibinfo {pages} {615} (\bibinfo {year} {1993})}\BibitemShut {NoStop}%
\bibitem [{\citenamefont {Li}(2002{\natexlab{a}})}]{Li:2002qx}%
  \BibitemOpen
  \bibfield  {author} {\bibinfo {author} {\bibfnamefont {B.-A.}\ \bibnamefont
  {Li}},\ }\href {\doibase 10.1103/PhysRevLett.88.192701} {\bibfield  {journal}
  {\bibinfo  {journal} {Phys. Rev. Lett.}\ }\textbf {\bibinfo {volume} {88}},\
  \bibinfo {pages} {192701} (\bibinfo {year} {2002}{\natexlab{a}})}\BibitemShut
  {NoStop}%
\bibitem [{\citenamefont {Li}(2002{\natexlab{b}})}]{Li:2002yda}%
  \BibitemOpen
  \bibfield  {author} {\bibinfo {author} {\bibfnamefont {B.-A.}\ \bibnamefont
  {Li}},\ }\href {\doibase 10.1016/S0375-9474(02)01018-7} {\bibfield  {journal}
  {\bibinfo  {journal} {Nucl. Phys.}\ }\textbf {\bibinfo {volume} {A708}},\
  \bibinfo {pages} {365} (\bibinfo {year} {2002}{\natexlab{b}})}\BibitemShut
  {NoStop}%
\bibitem [{\citenamefont {Li}(2015)}]{Li:2015hfa}%
  \BibitemOpen
  \bibfield  {author} {\bibinfo {author} {\bibfnamefont {B.-A.}\ \bibnamefont
  {Li}},\ }\href {\doibase 10.1103/PhysRevC.92.034603} {\bibfield  {journal}
  {\bibinfo  {journal} {Phys. Rev.}\ }\textbf {\bibinfo {volume} {C92}},\
  \bibinfo {pages} {034603} (\bibinfo {year} {2015})}\BibitemShut {NoStop}%
\bibitem [{\citenamefont {Thorne}(1977)}]{Thorne:1977}%
  \BibitemOpen
  \bibfield  {author} {\bibinfo {author} {\bibfnamefont {K.~S.}\ \bibnamefont
  {Thorne}},\ }\href@noop {} {\bibfield  {journal} {\bibinfo  {journal}
  {Astrophys. J.}\ }\textbf {\bibinfo {volume} {212}},\ \bibinfo {pages} {825}
  (\bibinfo {year} {1977})}\BibitemShut {NoStop}%
\bibitem [{\citenamefont {Brown}\ \emph {et~al.}(1998)\citenamefont {Brown},
  \citenamefont {Bildsten},\ and\ \citenamefont {Rutledge}}]{Brown:1998qv}%
  \BibitemOpen
  \bibfield  {author} {\bibinfo {author} {\bibfnamefont {E.~F.}\ \bibnamefont
  {Brown}}, \bibinfo {author} {\bibfnamefont {L.}~\bibnamefont {Bildsten}}, \
  and\ \bibinfo {author} {\bibfnamefont {R.~E.}\ \bibnamefont {Rutledge}},\
  }\href {\doibase 10.1086/311578} {\bibfield  {journal} {\bibinfo  {journal}
  {Astrophys. J.}\ }\textbf {\bibinfo {volume} {504}},\ \bibinfo {pages} {L95}
  (\bibinfo {year} {1998})}\BibitemShut {NoStop}%
\bibitem [{\citenamefont {Cackett}\ \emph {et~al.}(2008)\citenamefont
  {Cackett}, \citenamefont {Wijnands}, \citenamefont {Miller}, \citenamefont
  {Brown},\ and\ \citenamefont {Degenaar}}]{Cackett:2008tq}%
  \BibitemOpen
  \bibfield  {author} {\bibinfo {author} {\bibfnamefont {E.~M.}\ \bibnamefont
  {Cackett}}, \bibinfo {author} {\bibfnamefont {R.}~\bibnamefont {Wijnands}},
  \bibinfo {author} {\bibfnamefont {J.~M.}\ \bibnamefont {Miller}}, \bibinfo
  {author} {\bibfnamefont {E.~F.}\ \bibnamefont {Brown}}, \ and\ \bibinfo
  {author} {\bibfnamefont {N.}~\bibnamefont {Degenaar}},\ }\href {\doibase
  10.1086/593703} {\bibfield  {journal} {\bibinfo  {journal} {Astrophys. J.}\
  }\textbf {\bibinfo {volume} {687}},\ \bibinfo {pages} {L87} (\bibinfo {year}
  {2008})}\BibitemShut {NoStop}%
\bibitem [{\citenamefont {Page}\ and\ \citenamefont
  {Reddy}(2012)}]{Page:2012zt}%
  \BibitemOpen
  \bibfield  {author} {\bibinfo {author} {\bibfnamefont {D.}~\bibnamefont
  {Page}}\ and\ \bibinfo {author} {\bibfnamefont {S.}~\bibnamefont {Reddy}},\
  }\href@noop {} {\  (\bibinfo {year} {2012})},\ \Eprint
  {http://arxiv.org/abs/1201.5602} {arXiv:1201.5602 [nucl-th]} \BibitemShut
  {NoStop}%
\bibitem [{\citenamefont {{Brown}}(2015)}]{Brown:2015Code}%
  \BibitemOpen
  \bibfield  {author} {\bibinfo {author} {\bibfnamefont {E.~F.}\ \bibnamefont
  {{Brown}}},\ }\href@noop {} {\enquote {\bibinfo {title} {{dStar: Neutron star
  thermal evolution code}},}\ }\bibinfo {howpublished} {Astrophysics Source
  Code Library} (\bibinfo {year} {2015}),\ \Eprint
  {http://arxiv.org/abs/1505.034} {ascl:1505.034} \BibitemShut {NoStop}%
\bibitem [{\citenamefont {Deibel}\ \emph {et~al.}(2017)\citenamefont {Deibel},
  \citenamefont {Cumming}, \citenamefont {Brown},\ and\ \citenamefont
  {Reddy}}]{Deibel:2016vbc}%
  \BibitemOpen
  \bibfield  {author} {\bibinfo {author} {\bibfnamefont {A.}~\bibnamefont
  {Deibel}}, \bibinfo {author} {\bibfnamefont {A.}~\bibnamefont {Cumming}},
  \bibinfo {author} {\bibfnamefont {E.~F.}\ \bibnamefont {Brown}}, \ and\
  \bibinfo {author} {\bibfnamefont {S.}~\bibnamefont {Reddy}},\ }\href
  {\doibase 10.3847/1538-4357/aa6a19} {\bibfield  {journal} {\bibinfo
  {journal} {Astrophys. J.}\ }\textbf {\bibinfo {volume} {839}},\ \bibinfo
  {pages} {95} (\bibinfo {year} {2017})}\BibitemShut {NoStop}%
\bibitem [{\citenamefont {Brown}\ \emph {et~al.}(2018)\citenamefont {Brown},
  \citenamefont {Cumming}, \citenamefont {Fattoyev}, \citenamefont {Horowitz},
  \citenamefont {Page},\ and\ \citenamefont {Reddy}}]{Brown:2017gxd}%
  \BibitemOpen
  \bibfield  {author} {\bibinfo {author} {\bibfnamefont {E.~F.}\ \bibnamefont
  {Brown}}, \bibinfo {author} {\bibfnamefont {A.}~\bibnamefont {Cumming}},
  \bibinfo {author} {\bibfnamefont {F.~J.}\ \bibnamefont {Fattoyev}}, \bibinfo
  {author} {\bibfnamefont {C.~J.}\ \bibnamefont {Horowitz}}, \bibinfo {author}
  {\bibfnamefont {D.}~\bibnamefont {Page}}, \ and\ \bibinfo {author}
  {\bibfnamefont {S.}~\bibnamefont {Reddy}},\ }\href {\doibase
  10.1103/PhysRevLett.120.182701} {\bibfield  {journal} {\bibinfo  {journal}
  {Phys. Rev. Lett.}\ }\textbf {\bibinfo {volume} {120}},\ \bibinfo {pages}
  {182701} (\bibinfo {year} {2018})}\BibitemShut {NoStop}%
\bibitem [{\citenamefont {Cumming}\ \emph {et~al.}(2017)\citenamefont
  {Cumming}, \citenamefont {Brown}, \citenamefont {Fattoyev}, \citenamefont
  {Horowitz}, \citenamefont {Page},\ and\ \citenamefont
  {Reddy}}]{Cumming:2016weq}%
  \BibitemOpen
  \bibfield  {author} {\bibinfo {author} {\bibfnamefont {A.}~\bibnamefont
  {Cumming}}, \bibinfo {author} {\bibfnamefont {E.~F.}\ \bibnamefont {Brown}},
  \bibinfo {author} {\bibfnamefont {F.~J.}\ \bibnamefont {Fattoyev}}, \bibinfo
  {author} {\bibfnamefont {C.~J.}\ \bibnamefont {Horowitz}}, \bibinfo {author}
  {\bibfnamefont {D.}~\bibnamefont {Page}}, \ and\ \bibinfo {author}
  {\bibfnamefont {S.}~\bibnamefont {Reddy}},\ }\href {\doibase
  10.1103/PhysRevC.95.025806} {\bibfield  {journal} {\bibinfo  {journal} {Phys.
  Rev.}\ }\textbf {\bibinfo {volume} {C95}},\ \bibinfo {pages} {025806}
  (\bibinfo {year} {2017})}\BibitemShut {NoStop}%
\bibitem [{\citenamefont {Horowitz}\ \emph {et~al.}(2015)\citenamefont
  {Horowitz}, \citenamefont {Berry}, \citenamefont {Briggs}, \citenamefont
  {Caplan}, \citenamefont {Cumming},\ and\ \citenamefont
  {Schneider}}]{Horowitz:2014xca}%
  \BibitemOpen
  \bibfield  {author} {\bibinfo {author} {\bibfnamefont {C.~J.}\ \bibnamefont
  {Horowitz}}, \bibinfo {author} {\bibfnamefont {D.~K.}\ \bibnamefont {Berry}},
  \bibinfo {author} {\bibfnamefont {C.~M.}\ \bibnamefont {Briggs}}, \bibinfo
  {author} {\bibfnamefont {M.~E.}\ \bibnamefont {Caplan}}, \bibinfo {author}
  {\bibfnamefont {A.}~\bibnamefont {Cumming}}, \ and\ \bibinfo {author}
  {\bibfnamefont {A.~S.}\ \bibnamefont {Schneider}},\ }\href {\doibase
  10.1103/PhysRevLett.114.031102} {\bibfield  {journal} {\bibinfo  {journal}
  {Phys. Rev. Lett.}\ }\textbf {\bibinfo {volume} {114}},\ \bibinfo {pages}
  {031102} (\bibinfo {year} {2015})}\BibitemShut {NoStop}%
\bibitem [{\citenamefont {Schwenk}\ \emph {et~al.}(2003)\citenamefont
  {Schwenk}, \citenamefont {Friman},\ and\ \citenamefont
  {Brown}}]{Schwenk:2002fq}%
  \BibitemOpen
  \bibfield  {author} {\bibinfo {author} {\bibfnamefont {A.}~\bibnamefont
  {Schwenk}}, \bibinfo {author} {\bibfnamefont {B.}~\bibnamefont {Friman}}, \
  and\ \bibinfo {author} {\bibfnamefont {G.~E.}\ \bibnamefont {Brown}},\ }\href
  {\doibase 10.1016/S0375-9474(02)01290-3} {\bibfield  {journal} {\bibinfo
  {journal} {Nucl. Phys.}\ }\textbf {\bibinfo {volume} {A713}},\ \bibinfo
  {pages} {191} (\bibinfo {year} {2003})}\BibitemShut {NoStop}%
\bibitem [{\citenamefont {Gandolfi}\ \emph {et~al.}(2008)\citenamefont
  {Gandolfi}, \citenamefont {Illarionov}, \citenamefont {Fantoni},
  \citenamefont {Pederiva},\ and\ \citenamefont {Schmidt}}]{Gandolfi:2008id}%
  \BibitemOpen
  \bibfield  {author} {\bibinfo {author} {\bibfnamefont {S.}~\bibnamefont
  {Gandolfi}}, \bibinfo {author} {\bibfnamefont {A.~Y.}\ \bibnamefont
  {Illarionov}}, \bibinfo {author} {\bibfnamefont {S.}~\bibnamefont {Fantoni}},
  \bibinfo {author} {\bibfnamefont {F.}~\bibnamefont {Pederiva}}, \ and\
  \bibinfo {author} {\bibfnamefont {K.~E.}\ \bibnamefont {Schmidt}},\ }\href
  {\doibase 10.1103/PhysRevLett.101.132501} {\bibfield  {journal} {\bibinfo
  {journal} {Phys. Rev. Lett.}\ }\textbf {\bibinfo {volume} {101}},\ \bibinfo
  {pages} {132501} (\bibinfo {year} {2008})}\BibitemShut {NoStop}%
\bibitem [{\citenamefont {Cackett}\ \emph {et~al.}(2013)\citenamefont
  {Cackett}, \citenamefont {Brown}, \citenamefont {Cumming}, \citenamefont
  {Degenaar}, \citenamefont {Fridriksson}, \citenamefont {Homan}, \citenamefont
  {Miller},\ and\ \citenamefont {Wijnands}}]{Cackett:2013wva}%
  \BibitemOpen
  \bibfield  {author} {\bibinfo {author} {\bibfnamefont {E.~M.}\ \bibnamefont
  {Cackett}}, \bibinfo {author} {\bibfnamefont {E.~F.}\ \bibnamefont {Brown}},
  \bibinfo {author} {\bibfnamefont {A.}~\bibnamefont {Cumming}}, \bibinfo
  {author} {\bibfnamefont {N.}~\bibnamefont {Degenaar}}, \bibinfo {author}
  {\bibfnamefont {J.~K.}\ \bibnamefont {Fridriksson}}, \bibinfo {author}
  {\bibfnamefont {J.}~\bibnamefont {Homan}}, \bibinfo {author} {\bibfnamefont
  {J.~M.}\ \bibnamefont {Miller}}, \ and\ \bibinfo {author} {\bibfnamefont
  {R.}~\bibnamefont {Wijnands}},\ }\href {\doibase 10.1088/0004-637X/774/2/131}
  {\bibfield  {journal} {\bibinfo  {journal} {Astrophys. J.}\ }\textbf
  {\bibinfo {volume} {774}},\ \bibinfo {pages} {131} (\bibinfo {year}
  {2013})}\BibitemShut {NoStop}%
\bibitem [{\citenamefont {{Wijnands}}\ \emph {et~al.}(2003)\citenamefont
  {{Wijnands}}, \citenamefont {{Nowak}}, \citenamefont {{Miller}},
  \citenamefont {{Homan}}, \citenamefont {{Wachter}},\ and\ \citenamefont
  {{Lewin}}}]{wijnands2003}%
  \BibitemOpen
  \bibfield  {author} {\bibinfo {author} {\bibfnamefont {R.}~\bibnamefont
  {{Wijnands}}}, \bibinfo {author} {\bibfnamefont {M.}~\bibnamefont {{Nowak}}},
  \bibinfo {author} {\bibfnamefont {J.~M.}\ \bibnamefont {{Miller}}}, \bibinfo
  {author} {\bibfnamefont {J.}~\bibnamefont {{Homan}}}, \bibinfo {author}
  {\bibfnamefont {S.}~\bibnamefont {{Wachter}}}, \ and\ \bibinfo {author}
  {\bibfnamefont {W.~H.~G.}\ \bibnamefont {{Lewin}}},\ }\href {\doibase
  10.1086/377122} {\bibfield  {journal} {\bibinfo  {journal} {\apj}\ }\textbf
  {\bibinfo {volume} {594}},\ \bibinfo {pages} {952} (\bibinfo {year}
  {2003})}\BibitemShut {NoStop}%
\bibitem [{\citenamefont {{Wijnands}}\ \emph {et~al.}(2004)\citenamefont
  {{Wijnands}}, \citenamefont {{Homan}}, \citenamefont {{Miller}},\ and\
  \citenamefont {{Lewin}}}]{wijnands2004}%
  \BibitemOpen
  \bibfield  {author} {\bibinfo {author} {\bibfnamefont {R.}~\bibnamefont
  {{Wijnands}}}, \bibinfo {author} {\bibfnamefont {J.}~\bibnamefont {{Homan}}},
  \bibinfo {author} {\bibfnamefont {J.~M.}\ \bibnamefont {{Miller}}}, \ and\
  \bibinfo {author} {\bibfnamefont {W.~H.~G.}\ \bibnamefont {{Lewin}}},\ }\href
  {\doibase 10.1086/421081} {\bibfield  {journal} {\bibinfo  {journal}
  {Astrophys. J. Lett.}\ }\textbf {\bibinfo {volume} {606}},\ \bibinfo {pages}
  {L61} (\bibinfo {year} {2004})}\BibitemShut {NoStop}%
\bibitem [{\citenamefont {{Negoro}}\ \emph {et~al.}(2015)\citenamefont
  {{Negoro}}, \citenamefont {{Furuya}}, \citenamefont {{Ueno}}, \citenamefont
  {{Tomida}}, \citenamefont {{Nakahira}}, \citenamefont {{Kimura}},
  \citenamefont {{Ishikawa}}, \citenamefont {{Nakagawa}}, \citenamefont
  {{Mihara}}, \citenamefont {{Sugizaki}}, \citenamefont {{Serino}},
  \citenamefont {{Shidatsu}}, \citenamefont {{Sugimoto}}, \citenamefont
  {{Takagi}}, \citenamefont {{Matsuoka}}, \citenamefont {{Kawai}},
  \citenamefont {{Tachibana}}, \citenamefont {{Yoshii}}, \citenamefont
  {{Yoshida}}, \citenamefont {{Sakamoto}}, \citenamefont {{Kawakubo}},
  \citenamefont {{Ohtsuki}}, \citenamefont {{Tsunemi}}, \citenamefont
  {{Imatani}}, \citenamefont {{Nakajima}}, \citenamefont {{Masumitsu}},
  \citenamefont {{Tanaka}}, \citenamefont {{Ueda}}, \citenamefont {{Kawamuro}},
  \citenamefont {{Hori}}, \citenamefont {{Tsuboi}}, \citenamefont {{Kanetou}},
  \citenamefont {{Yamauchi}}, \citenamefont {{Itoh}}, \citenamefont
  {{Yamaoka}},\ and\ \citenamefont {{Morii}}}]{negoro2015}%
  \BibitemOpen
  \bibfield  {author} {\bibinfo {author} {\bibfnamefont {H.}~\bibnamefont
  {{Negoro}}}, \bibinfo {author} {\bibfnamefont {K.}~\bibnamefont {{Furuya}}},
  \bibinfo {author} {\bibfnamefont {S.}~\bibnamefont {{Ueno}}}, \bibinfo
  {author} {\bibfnamefont {H.}~\bibnamefont {{Tomida}}}, \bibinfo {author}
  {\bibfnamefont {S.}~\bibnamefont {{Nakahira}}}, \bibinfo {author}
  {\bibfnamefont {M.}~\bibnamefont {{Kimura}}}, \bibinfo {author}
  {\bibfnamefont {M.}~\bibnamefont {{Ishikawa}}}, \bibinfo {author}
  {\bibfnamefont {Y.~E.}\ \bibnamefont {{Nakagawa}}}, \bibinfo {author}
  {\bibfnamefont {T.}~\bibnamefont {{Mihara}}}, \bibinfo {author}
  {\bibfnamefont {M.}~\bibnamefont {{Sugizaki}}}, \bibinfo {author}
  {\bibfnamefont {M.}~\bibnamefont {{Serino}}}, \bibinfo {author}
  {\bibfnamefont {M.}~\bibnamefont {{Shidatsu}}}, \bibinfo {author}
  {\bibfnamefont {J.}~\bibnamefont {{Sugimoto}}}, \bibinfo {author}
  {\bibfnamefont {T.}~\bibnamefont {{Takagi}}}, \bibinfo {author}
  {\bibfnamefont {M.}~\bibnamefont {{Matsuoka}}}, \bibinfo {author}
  {\bibfnamefont {N.}~\bibnamefont {{Kawai}}}, \bibinfo {author} {\bibfnamefont
  {Y.}~\bibnamefont {{Tachibana}}}, \bibinfo {author} {\bibfnamefont
  {T.}~\bibnamefont {{Yoshii}}}, \bibinfo {author} {\bibfnamefont
  {A.}~\bibnamefont {{Yoshida}}}, \bibinfo {author} {\bibfnamefont
  {T.}~\bibnamefont {{Sakamoto}}}, \bibinfo {author} {\bibfnamefont
  {Y.}~\bibnamefont {{Kawakubo}}}, \bibinfo {author} {\bibfnamefont
  {H.}~\bibnamefont {{Ohtsuki}}}, \bibinfo {author} {\bibfnamefont
  {H.}~\bibnamefont {{Tsunemi}}}, \bibinfo {author} {\bibfnamefont
  {R.}~\bibnamefont {{Imatani}}}, \bibinfo {author} {\bibfnamefont
  {M.}~\bibnamefont {{Nakajima}}}, \bibinfo {author} {\bibfnamefont
  {T.}~\bibnamefont {{Masumitsu}}}, \bibinfo {author} {\bibfnamefont
  {K.}~\bibnamefont {{Tanaka}}}, \bibinfo {author} {\bibfnamefont
  {Y.}~\bibnamefont {{Ueda}}}, \bibinfo {author} {\bibfnamefont
  {T.}~\bibnamefont {{Kawamuro}}}, \bibinfo {author} {\bibfnamefont
  {T.}~\bibnamefont {{Hori}}}, \bibinfo {author} {\bibfnamefont
  {Y.}~\bibnamefont {{Tsuboi}}}, \bibinfo {author} {\bibfnamefont
  {S.}~\bibnamefont {{Kanetou}}}, \bibinfo {author} {\bibfnamefont
  {M.}~\bibnamefont {{Yamauchi}}}, \bibinfo {author} {\bibfnamefont
  {D.}~\bibnamefont {{Itoh}}}, \bibinfo {author} {\bibfnamefont
  {K.}~\bibnamefont {{Yamaoka}}}, \ and\ \bibinfo {author} {\bibfnamefont
  {M.}~\bibnamefont {{Morii}}},\ }\href@noop {} {\bibfield  {journal} {\bibinfo
   {journal} {The Astronomer's Telegram}\ }\textbf {\bibinfo {volume} {7943}},\
  \bibinfo {pages} {1} (\bibinfo {year} {2015})}\BibitemShut {NoStop}%
\bibitem [{\citenamefont {Zel'dovich}\ and\ \citenamefont
  {{Shakura}}(1969)}]{Zeldovich:1969}%
  \BibitemOpen
  \bibfield  {author} {\bibinfo {author} {\bibfnamefont {Y.~B.}\ \bibnamefont
  {Zel'dovich}}\ and\ \bibinfo {author} {\bibfnamefont {N.~I.}\ \bibnamefont
  {{Shakura}}},\ }\href@noop {} {\bibfield  {journal} {\bibinfo  {journal}
  {Soviet Astronomy}\ }\textbf {\bibinfo {volume} {13}},\ \bibinfo {pages}
  {175} (\bibinfo {year} {1969})}\BibitemShut {NoStop}%
\bibitem [{\citenamefont {Shapiro}\ and\ \citenamefont
  {{Salpeter}}(1975)}]{Shapiro:1975}%
  \BibitemOpen
  \bibfield  {author} {\bibinfo {author} {\bibfnamefont {S.~L.}\ \bibnamefont
  {Shapiro}}\ and\ \bibinfo {author} {\bibfnamefont {E.~E.}\ \bibnamefont
  {{Salpeter}}},\ }\href@noop {} {\bibfield  {journal} {\bibinfo  {journal}
  {Astrophys. J.}\ }\textbf {\bibinfo {volume} {198}},\ \bibinfo {pages} {671}
  (\bibinfo {year} {1975})}\BibitemShut {NoStop}%
\bibitem [{\citenamefont {Bildsten}\ \emph {et~al.}(2003)\citenamefont
  {Bildsten}, \citenamefont {Chang},\ and\ \citenamefont
  {Paerels}}]{Bildsten:2003ApJ}%
  \BibitemOpen
  \bibfield  {author} {\bibinfo {author} {\bibfnamefont {L.}~\bibnamefont
  {Bildsten}}, \bibinfo {author} {\bibfnamefont {P.}~\bibnamefont {Chang}}, \
  and\ \bibinfo {author} {\bibfnamefont {F.}~\bibnamefont {Paerels}},\ }\href
  {\doibase 10.1086/377066} {\bibfield  {journal} {\bibinfo  {journal}
  {Astrophys. J.}\ }\textbf {\bibinfo {volume} {591}},\ \bibinfo {pages} {L29}
  (\bibinfo {year} {2003})}\BibitemShut {NoStop}%
\end{thebibliography}%
\end{document}